\documentclass[submission, Phys]{SciPost}

\usepackage[utf8]{inputenc} % input encodings (allow UTF-8 input)
\usepackage[T1]{fontenc} 	% selecting font encodings (use 8-bit T1 fonts)
\usepackage[english]{babel} % multilingual support (English language/hyphenation)

\usepackage[bitstream-charter]{mathdesign}
\usepackage{geometry} 		% flexible and complete interface to document dimensions
\usepackage{amsmath} 		% math package (American Mathematical Society)
\usepackage{mathtools} 		% math package (fixes various deficiencies of amsmath)
\usepackage{float} 			% floating objects such as figures and tables
\usepackage{graphicx} 		% enhanced support for graphics
\usepackage{tabularx} 		% tabulars with adjustable-width columns
\usepackage{booktabs} 		% professional-quality tables
\usepackage{color, xcolor} 	% foreground and background colour management
\usepackage{pdfpages} 		% inclusion of external multi-page PDF documents
\usepackage{extarrows} 		% extra arrows beyond those provided in amsmath
\usepackage{multirow} 		% create tabular cells spanning multiple rows
\usepackage{multicol} 		% intermix single and multiple columns
\usepackage{enumitem} 		% control layout of itemize, enumerate, description
\usepackage{xspace} 		% define commands that appear not to eat spaces
\usepackage{stackrel} 		% enhancement to the \stackrel command
\usepackage{tikz} 			% create PostScript and PDF graphics
\usepackage{braket} 		% Dirac bra-ket notation
\usepackage{bm} 			% access bold symbols in maths mode
\usepackage{tensor} 		% typeset tensors (tensor-style super- and subscripts)
\usepackage{slashed} 		% slash through characters (Feynman slash notation)
\usepackage{siunitx} 		% SI units package (typesetting values with units)
\usepackage{lastpage} 		% reference last page
\usepackage{cite} 			% improved citation handling
\usepackage[normalem]{ulem} % package for underlining
\usepackage{fontawesome} 	% access to web-related icons
\usepackage{tocloft} 		% control over the typography of the TOC, LOF, and LOT
\usepackage{titlesec} 		% interface to sectioning commands (various title styles)
\usepackage{doi} 			% create correct hyperlinks for DOI numbers
\usepackage{hyperref} 		% hypertext links (handel cross-referencing commands)
\usepackage[most]{tcolorbox} 					% coloured and framed text boxes
\usepackage[nameinlink, capitalize]{cleveref} 	% intelligent cross-referencing
\usepackage[nottoc, notlot, notlof]{tocbibind} 	% add references/index/contents to TOC
\usepackage[ruled, vlined]{algorithm2e} 		% floating algorithm environment
\usepackage{makecell}
\usepackage[makeroom]{cancel}
\usepackage{feynmf}
\usepackage{sepfootnotes}
\makeatletter
\def\BState{\State\hskip-\ALG@thistlm}
\makeatother

\makeatletter
\@ifundefined{pdfoutput}{}{\DeclareGraphicsRule{*}{mps}{*}{}}
\makeatother

\makeatletter
\DeclareRobustCommand*{\bfseries}{%
   \not@math@alphabet\bfseries\mathbf
   \fontseries\bfdefault\selectfont
   \boldmath
}
\makeatother

\hypersetup{
	pdftitle={NN_Amp_CS},
	pdfauthor={Beccatini et al.},
	colorlinks=true, 			% false: boxed links, true: colored links
	linkcolor={red!50!black}, 	% color of internal links (sections, pages, etc.)
	citecolor={blue!50!black}, 	% color of citation links (links to bibliography)
	urlcolor={blue!80!black} 	% color of URL links (external links)
} 
\DeclareSymbolFont{usualmathcal}{OMS}{cmsy}{m}{n}
\DeclareSymbolFontAlphabet{\mathcal}{usualmathcal}

\SetArgSty{textnormal}
\SetKwComment{Comment}{{\small\#}~}{}
\SetCommentSty{mycommfont}

\setitemize{itemsep=0pt, parsep=0pt} 				% adjust itemize environment
\setenumerate{itemsep=0pt, parsep=0pt} 				% adjust enumerate environment
\setitemize{itemsep=2pt,topsep=2pt,parsep=0pt,partopsep=0pt,leftmargin=*}
\setenumerate{itemsep=0pt,topsep=2pt,parsep=0pt,partopsep=0pt,labelindent=3pt,leftmargin=*}
\usepackage{amsmath}
 
\usepackage{amsthm} 		% math package (typesetting theorems using AMS style)
\theoremstyle{definition}

\definecolor{red_cb}{HTML}{e41a1c}
\definecolor{blue_cb}{HTML}{377eb8}
\definecolor{green_cb}{HTML}{4daf4a}
\definecolor{purple_cb}{HTML}{984ea3}
\definecolor{orange_cb}{HTML}{ff7f00}

\definecolor{EmeraldGreen}{HTML}{1ea78d}
\definecolor{EnglishRed}{HTML}{b02427}
\hypersetup{colorlinks=true,urlcolor=EmeraldGreen,citecolor=EmeraldGreen,linkcolor=EnglishRed}

\newcommand{\ie}{\text{i.e.}\;}

\newcommand{\mwith}{\text{with}}
\newcommand{\mand}{\text{and}}

\newcommand{\Langle}{\bigl\langle}
\newcommand{\Rangle}{\bigr\rangle}

\def\d{\mathrm{d}}

\newcommand\one{\leavevmode\hbox{\small1\normalsize\kern-.33em1}}
\newcommand{\loss}{\mathcal{L}} 	% loss value

\newcommand{\mg}{\textsc{MG5aMC}\xspace}
\newcommand{\pythia}{\textsc{Pythia8}\xspace}

\newcommand{\sherpa}{\textsc{Sherpa}\xspace}
\newcommand{\herwig}{\textsc{Herwig}\xspace}
\newcommand{\madnis}{\textsc{MadNIS}\xspace}

\newcommand{\arXiv}[2][]{%
	\ifthenelse{\equal{#1}{}}%
	{\href{http://arxiv.org/abs/#2}{arXiv:#2}}%
	{\href{http://arxiv.org/abs/#2}{arXiv:#2~[#1]}}}

\def\slashchar#1{\setbox0=\hbox{$#1$}           % set a box for #1
   \dimen0=\wd0                                 % and get its size
   \setbox1=\hbox{/} \dimen1=\wd1               % get size of /
   \ifdim\dimen0>\dimen1                        % #1 is bigger
      \rlap{\hbox to \dimen0{\hfil/\hfil}}      % so center / in box
      #1                                        % and print #1
   \else                                        % / is bigger
      \rlap{\hbox to \dimen1{\hfil$#1$\hfil}}   % so center #1
      /                                         % and print /
   \fi}

\newcommand{\tikznode}[2]{%
\ifmmode%
\tikz[remember picture,baseline=(#1.base),inner sep=0pt] \node (#1) {$#2$};%
\else
\tikz[remember picture,baseline=(#1.base),inner sep=0pt] \node (#1) {#2};%
\fi}

\def\mathswitchr#1{\relax\ifmmode{\mathrm{#1}}\else$\mathrm{#1}$\xspace\fi}
\def\mathswitch#1{\relax\ifmmode#1\else$#1$\xspace\fi}

\newcommand{\PZ}{\mathswitchr Z}

\newcommand{\Pg}{\mathswitchr g}

\newcommand{\Pd}{\mathswitchr d}

\newcommand{\Pdbar}{\mathswitchr{\bar d}}

\graphicspath{{./figures/}}

\begin{document}

% article title
\begin{center}
{\Large\textbf{
Amplitude Surrogates for Multi-Jet Processes}}
\end{center}

\begin{center}
  Luca Beccatini\textsuperscript{1,2,3},
  Fabio Maltoni\textsuperscript{1,2,3,4},
  Olivier Mattelaer\textsuperscript{1}, and
  Ramon Winterhalder\textsuperscript{5}
\end{center}

% affiliations (format: institute, city, country)
\begin{center}
  {\bf 1} CP3, Universit\'e catholique de Louvain, Louvain-la-Neuve, Belgium \\
  {\bf 2} Dipartimento di Fisica e Astronomia, Universit\`a di Bologna, Italy \\
  {\bf 3} INFN, Sezione di Bologna, Bologna, Italy\\
  {\bf 4} European Organisation for Nuclear Research (CERN), Geneva, Switzerland\\
  {\bf 5} TIFLab, Universit\`a degli Studi di Milano \& INFN Sezione di Milano, Milano, Italy
\end{center}

\begin{center}
    \today
\end{center}

\vspace{-1cm}

\section*{Abstract}
{\bf Accurate and efficient amplitude predictions are essential for precision studies of multi-jet processes at the LHC. We introduce a novel neural network architecture that predicts multi-jet amplitudes by leveraging the Catani–Seymour factorization scheme and related lower-jet amplitudes, requiring the network to learn only a correction factor. This hybrid approach combines theoretical factorization with a data-driven ansatz, enabling fast and scalable amplitude predictions. Our networks also estimate the accuracy of each prediction, allowing us to selectively use results that meet a predefined accuracy threshold. In the context of leading-order event generation, this approach achieves speed-up factors of up to 20 while maintaining all observables at the percent-level accuracy.}

% 

% include a table of contents (optional)
\vspace{10pt}
\noindent\rule{\linewidth}{1pt}
\tableofcontents\thispagestyle{fancy}
\noindent\rule{\linewidth}{1pt}
\vspace{10pt}

\clearpage

%%%%%%%%%%%%%%%%%%%%%%%%%%%%%%%%%%%%%%%%%%%%%%%%%%%%
\section{Introduction}
\label{sec:intro}

Accurate and efficient predictions of scattering amplitudes are essential for precision studies in high-energy physics, in particular for multi-jet processes at hadron colliders~\cite{Campbell:2022qmc}. At the LHC and its upcoming high-luminosity phase, increasingly differential measurements and complex final states demand Monte-Carlo simulations that combine large event samples with high perturbative accuracy. Multi-purpose event generators such as \pythia~\cite{Sjostrand:2014zea}, \sherpa~\cite{Sherpa:2019gpd}, \herwig~\cite{Bellm:2017bvx}, and \mg~\cite{Alwall:2014hca} provide the backbone of this theoretical infrastructure by interfacing fixed-order matrix elements with parton showers, hadronization, and detector simulation.

Despite many algorithmic advances, evaluating scattering amplitudes remains a significant computational bottleneck in event generation. The cost of computing tree-level amplitudes grows rapidly with final-state multiplicity, reflecting both the combinatorial growth of Feynman diagrams and the increasing complexity of phase-space integration. While more advanced techniques~\cite{Bolinder:2025gbj} substantially mitigate the na\"ive factorial scaling, high-multiplicity matrix-element evaluations still account for a non-negligible fraction of the total runtime in modern simulation workflows \cite{HEPSoftwareFoundation:2017ggl}.

A broad range of strategies has been developed to address this challenge. On the one hand, significant speed-ups have been achieved through hardware acceleration. In particular, GPU based computation, pioneered with MadGraph4 \cite{Alwall:2007st} more than 10 years ago \cite{Hagiwara:2010ujr,Hagiwara:2010oca, Hagiwara:2013oka},  are now production ready~\cite{Hageboeck:2023blb,Wettersten:2025hrb,Valassi:2025xfn,Hagebock:2025jyk}. Complementary developments, such as the \textsc{Pepper} framework~\cite{Bothmann:2023gew}, provide process specific GPU-accelerated event generation, including both fast amplitude evaluation and simplified integration routines.

In parallel, modern machine-learning techniques~\cite{Plehn:2022ftl} have emerged as a powerful tool to accelerate event generation~\cite{Butter:2022rso}. Considerable progress has been made in improving phase-space sampling through importance sampling, from early neural network approaches~\cite{Bendavid:2017zhk,Klimek:2018mza,Chen:2020nfb} to normalizing flows~\cite{Gao:2020vdv,Bothmann:2020ywa,Gao:2020zvv,Winterhalder:2021ngy,Bothmann:2025lwg,Janssen:2025zke} and dedicated frameworks such as \madnis~\cite{Heimel:2022wyj,Heimel:2023ngj,Heimel:2024wph}. While these methods can dramatically improve integration efficiency and unweighting, they typically leave the matrix-element evaluation itself untouched. 

A more radical direction is to replace expensive first-principles amplitude calculations by fast machine-learning surrogates. Early work demonstrated the feasibility of learning scattering amplitudes or matrix-element weights directly from data using neural networks~\cite{Bishara:2019iwh,Badger:2020uow,Aylett-Bullock:2021hmo, Maitre:2021uaa, Maitre:2023dqz, Breso:2024jlt}. Since then, a wide range of increasingly structured approaches has been explored. These include 
combinations with multi-stage unweighting strategies~\cite{Danziger:2021eeg,Janssen:2023ahv,Herrmann:2025nnz, Villadamigo:2025our}, Lorentz-equivariant architectures~\cite{Spinner:2024hjm,Brehmer:2024yqw,Favaro:2025pgz}, and the incorporation of learned uncertainty estimates~\cite{Badger:2022hwf,Bahl:2024gyt,Bahl:2025xvx}.
An important development in this context is the use of physics-informed surrogate targets, in which neural networks learn ratios or correction factors with respect to analytically motivated approximations rather than absolute amplitudes or weights. In particular, the works of Ref.~\cite{Maitre:2021uaa,Maitre:2023dqz} already exploit factorization properties of QCD amplitudes to construct surrogates with reduced dynamic range, significantly simplifying the learning task.

In this work, we pursue a closely related but distinct strategy that embeds QCD factorization more directly into the surrogate construction. Rather than correcting an analytic approximation at fixed multiplicity, we exploit the universal factorization structure of QCD amplitudes to relate an $n$-parton configuration to an exact reduced $(n-1)$-parton process. Our approach is rooted in the Catani–Seymour dipole formalism~\cite{Catani:1996vz,Catani:2002hc}, which provides an exact kinematic mapping between resolved and unresolved configurations. Starting from amplitudes for related processes with fewer final-state partons, we construct a factorized approximation of the full multi-jet amplitude and train a neural network to learn a correction factor that accounts for the residual difference to the exact result. This hybrid approach combines the strengths of theory-driven factorization and data-driven learning. By construction, a large fraction of the kinematic complexity is absorbed into the reduced-multiplicity amplitude. Therefore, the neural network only needs to learn a smoother, lower-variance function that is free of explicit singularities. This leads to improved accuracy and robustness, particularly for high-multiplicity final states and in regions of phase space that are challenging for purely data-driven surrogates.

We apply this method to multi-jet production in association with a $\PZ$ boson at leading order (LO) and show that factorization-based surrogates achieve high per-event accuracy across the full phase space. We further demonstrate how learned uncertainty estimates can be used to deploy the surrogate selectively within event generation, ensuring controlled numerical precision while achieving substantial speed-ups relative to standard matrix-element evaluations. Our approach complements advances in GPU acceleration and phase-space sampling and provides a viable path towards scalable event generation for high-multiplicity final states at future collider experiments as well as for the success of the high-luminosity runs~\cite{HEPSoftwareFoundation:2017ggl}.

The paper is organized as follows: 
In Sec.~\ref{sec:factorization}, we introduce the factorization ansatz underlying our surrogate construction, based on the Catani–Seymour factorization formalism.
In Sec.~\ref{sec:ml}, we describe the neural-network architecture and training strategy. In Sec.~\ref{sec:results}, we apply the method to $Z$+multi-jet production at leading order and assess the surrogate accuracy. In Sec.~\ref{sec:speedup}, we demonstrate its use within event generation and quantify the achievable speed-up. We conclude in Sec.~\ref{sec:outlook}.

%%%%%%%%%%%%%%%%%%%%%%%%%%%%%%%%%%%%%%%%%%%%%%%%%%%%
\section{Splitting kernel factorization}
\label{sec:factorization}

In this section, we review the factorization used as an ansatz for our neural-network surrogates. These formulas are based on the Catani–Seymour (CS) factorization, which accurately reproduces the expected behavior of QCD amplitudes in both the soft and collinear limits~\cite{Catani:1996vz}. They ensure exact momentum conservation by redefining the momenta of the emitter and spectator particles, allowing the dipole terms to be applied even outside the strict soft or collinear regimes without violating momentum conservation or producing off-shell emitters. This section focuses on the conceptual aspects, with the full-fledged formulas provided in App.~\ref{sec:CS_formulas}.

%%%%%%%%%%%%%%%%%%%%%%%%%%%%%%%%%%%%%%%%%%%%%%%%%%%%
\subsection{Factorization ansatz}

Given a scattering process with $n$ final-state particles and at least one quark/gluon, we define the ensemble of reduced processes as the set of all $n-1$-particle final states obtained by absorbing one of the original quark/gluon into a pair of emitter and spectator particles. This is schematically expressed as:
\begin{align}
(p_a, p_b \to p_1, \dots, p_i, p_j, p_k, \dots, p_{n})
\Longrightarrow
\left\{ (p_a, p_b \to p_1, \dots, \tilde{p}_{ij}, \tilde{p}_k, \dots, p_{n-1}) \right\}_{(i,j,k)}
\label{eq:FaCS_mom}
\end{align}
Here, $p_i, p_j, p_k$ denote the four-momenta of the emitter, emitted parton (quark/gluon), and spectator, respectively. The momenta $\tilde{p}_{ij}, \tilde{p}_k$ are redefined according to the CS prescription (See App.~\ref{sec:CS_formulas}, Eq.\eqref{eq:cs_remap}). The notation $(i,j,k)$ runs over all valid combinations of emitter, emitted, and spectator particles.

We define the matrix-element squared, summed and/or averaged over colours and spins, of a process with $n$ final-state particles as 
\begin{align}
\label{eq:def:amplitude}
    A_n \equiv \langle |\mathcal{M}(p_a, p_b \to p_1, \dots, p_n)|^2\rangle\;.
\end{align}
which we will simply refer to as the \emph{amplitude} in the following. Inspired by the CS formalism, we propose a general ansatz to approximate the amplitude of the full $n$-particle process, $A_{n}$, in terms of a corresponding reduced $(n-1)$-particle  amplitude, $A_{n-1}$.
This factorized form is given by
\begin{align}
    A_{n} \approx A_{n-1} \cdot F_{ij,k}^r\;.
    \label{eq:FaCS_approx}
\end{align}
The reduced amplitude $A_{n-1}$ depends on the set of reduced momenta, as shown in Eq.\eqref{eq:FaCS_mom}, for given particles $i, j, k$.
The function $F_{ij,k}^r$ represents an approximate splitting kernel, which depends on the kinematics of particles $i, j, k$, as well as the nature of the emitted radiation $r$. 

The factorization ansatz introduced in Eq.~\ref{eq:FaCS_approx} is derived from the Catani–Seymour dipole formalism, subject to two simplifying assumptions. First, we neglect spin–helicity correlations by replacing the spin-dependent dipole kernel with its spin-averaged counterpart. Second, we adopt the leading-colour approximation, reducing the full colour-correlated structure to its dominant contribution. Under these assumptions, the dipole factorization takes a simplified scalar form, which serves as the foundation of our ansatz. In Appendix~\ref{sec:CS_formulas}, we present the Catani–Seymour dipole formalism for both initial- and final-state radiation with a final-state spectator, and we compare our factorization ansatz directly to the corresponding exact dipole expressions.

The choice of the spectator particle in the dipole construction is, in principle, arbitrary. For a fixed selection of emitter and emitted parton, different choices of the spectator lead to distinct reduced kinematics, due to the momentum redefinition applied to both the emitter and the spectator. In our study, we observed that restricting the choice of spectator to final-state jets improves the performance of the neural network, leading to better accuracy. Therefore, for both practical and methodological reasons, we restrict our analysis in this paper to cases where the spectator is chosen as the most energetic final-state jet, in the laboratory frame, that is neither the emitter nor the emitted parton.

\subsection{Radiation type and rank}

Usually, a process admits several reduced processes. Equation~\eqref{eq:FaCS_approx} requires only one reduced process, implying that we must choose both the type of radiation and the specific emitter–jet combination to use in our ansatz. 

As the CS factorization accurately reproduces the QCD behavior in the soft and collinear limits, Eq.\eqref{eq:FaCS_approx} provides higher accuracy for events near these singular regions. To achieve the best possible approximation, we rank all candidate reduced processes for a given event according to the degree of singularity of their corresponding radiation, and select the most singular ones.

For example, consider the process $\Pd \Pdbar \to \PZ \Pg_1 \Pg_2 \Pg_3 \Pg_4$. If we look to the possible gluon splitting of the form $g_f \to g_f + g_f$, there are six distinct combinations where the emitted particles $p_i, p_j$ are
\begin{align}
\left\{g_i, \, g_j\right\}_{(i,j) = (1, 2), (1, 3), (1, 4), (2, 3), (2, 4), (3, 4)}
\end{align}
excluding symmetric configurations. By computing the scalar product $p_i \, p_j$ for each pair, we can rank the radiations from the most singular (smallest $p_i \, p_j$) to the least singular (largest $p_i \, p_j$). 
We define the rank of a radiation as its order in a scale of the most singular radiations of the same type. The first rank corresponds to the most singular radiations of the corresponding type, the second rank to the second most singular, and so on. 

Additionally, a given process may contain several distinct radiation channels, determined by the identities of the emitter and emitted particles. Since each radiation type is associated with a different splitting kernel, we introduce the following radiation labels to distinguish them
\begin{itemize}
\item FG1: Final state gluon radiation ($g_f \to g_f + g_f$);
\item FQ1: Final state quark radiation ($q_f \to q_f + g_f$);
\item IG1: Initial state gluon radiation ($g_i \to g_i + g_f$);
\item IQ1: Initial state quark radiation ($q_i \to q_i + g_f$).
\end{itemize}
The presence of the index 1 inside the radiation labels is related to the radiation rank and can be replaced by another digit accordingly. This means that FG2 corresponds to the second rank radiation within the final state gluon radiation. 

%------------------------------
\begin{figure}[tb!]
    \includegraphics[width=0.49\linewidth]{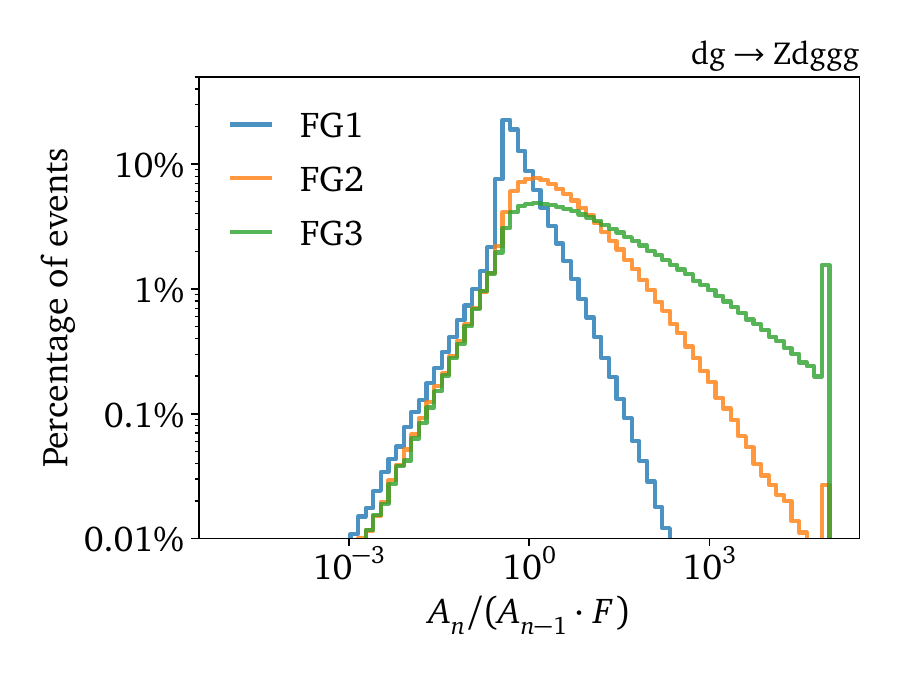}
    \includegraphics[width=0.49\linewidth]{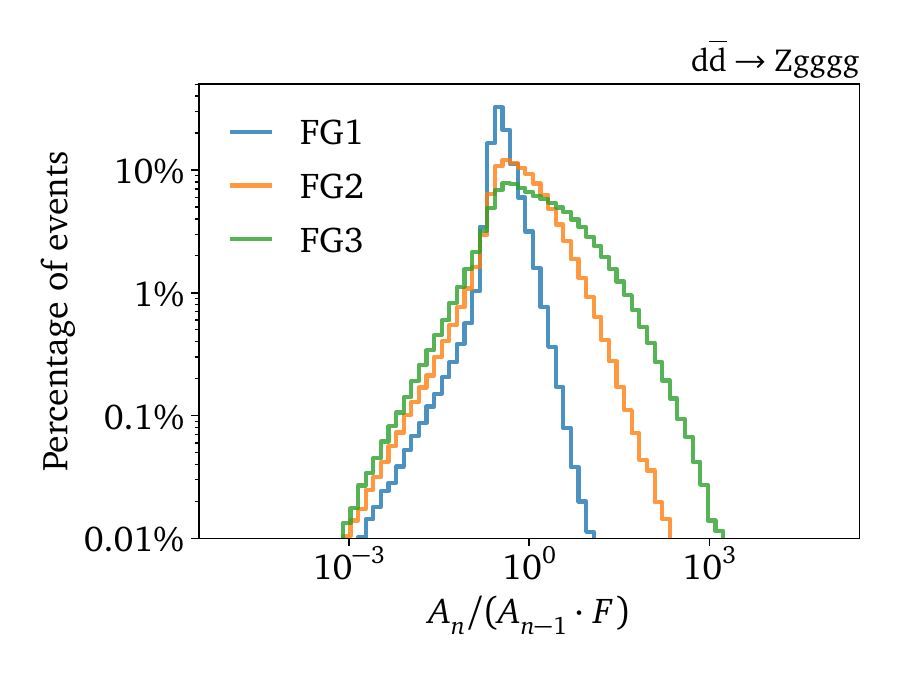}
    \caption{Approximation quality comparison between FG1, FG2, and FG3 radiations for $\Pd \Pg \to \PZ \Pd \Pg \Pg \Pg$ (left) and for $\Pd \Pdbar \to \PZ \Pg \Pg \Pg \Pg$ (right), obtained from unweighted samples.}
    \label{fig:Plot_Fact_Approx_rank}
\end{figure}
%------------------------------

In Fig.~\ref{fig:Plot_Fact_Approx_rank}, we illustrate the factorization quality  (ratio between the target and the surrogate) of the first three ranks for $g_f \to g_f + g_f$ splittings for $\Pd \Pg \to \PZ \Pd \Pg \Pg \Pg$ and $\Pd \Pdbar \to \PZ \Pg \Pg \Pg \Pg$. Since, in our ansatz, the surrogate output will eventually be rescaled by a neural-network correction factor, the absolute value of the ratio is not particularly relevant. What matters instead is the shape of this distribution. A narrower and less dispersed distribution implies that the correction factor the network must learn is simpler and more stable, leading to a more efficient and accurate training procedure.

As expected, the most singular radiation yields the highest quality, while radiations with lower singularity exhibit broader distributions. Moreover, we notice that $\Pd \Pdbar \to \PZ \Pg \Pg \Pg \Pg$ have better approximations than $\Pd \Pg \to \PZ \Pd \Pg \Pg \Pg$. In fact, having 4 gluons in the final state, it has more radiations of the same type ($g_f \to g_f + g_f$) in respect to the other process, increasing the probability of finding a more singular pair, and so, a better factorization for the event. 
This behaviour confirms that the approximation quality is significantly higher for singular configurations (soft and/or collinear emissions) compared to the rest of the phase space.

\begin{figure}[tb!]
    \includegraphics[width=0.49\linewidth]{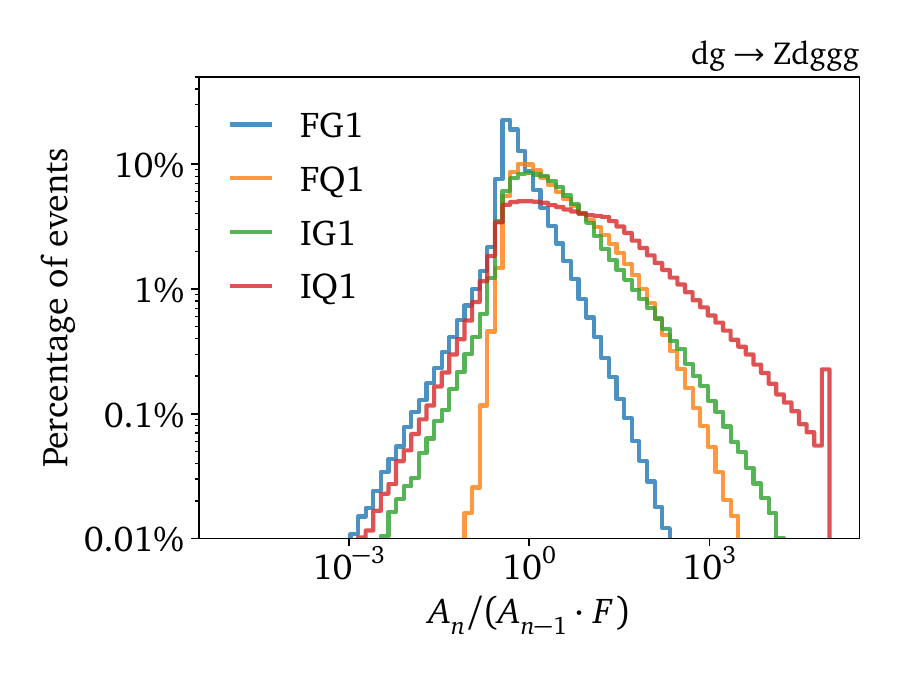}
    \includegraphics[width=0.49\linewidth]{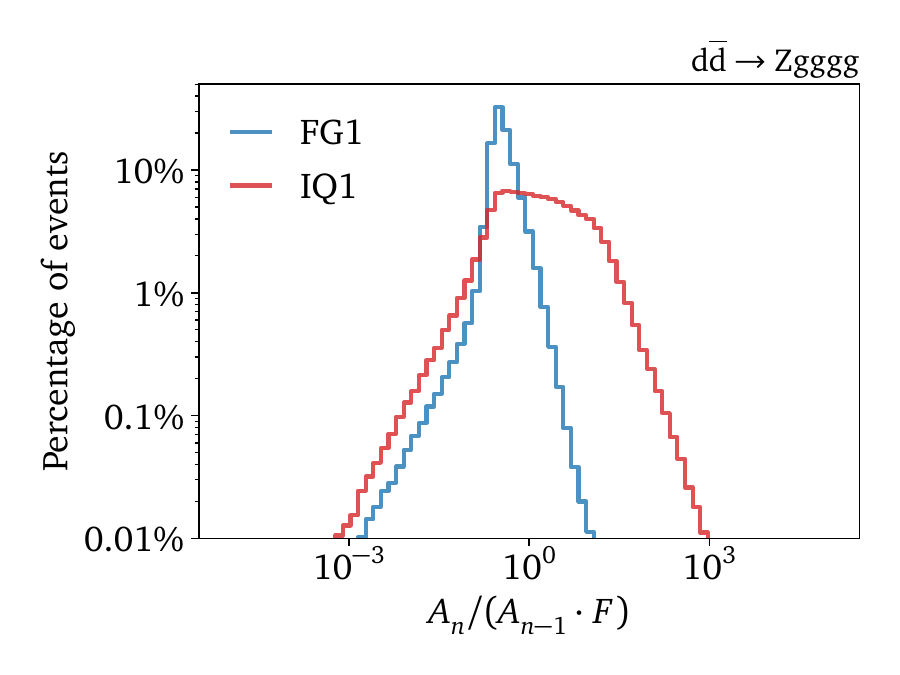}
    \caption{Approximation quality comparison between different radiation types: FG1, FQ1, IG1 and IQ1 radiations for $\Pd \Pg \to \PZ \Pd \Pg \Pg \Pg$ (left), and FG1, and IQ1 radiations for $\Pd \Pdbar \to \PZ \Pg \Pg \Pg \Pg$ (right), obtained from unweighted samples.} 
    \label{fig:Plot_Fact_Approx_type}
\end{figure}

In Fig.~\ref{fig:Plot_Fact_Approx_type}, we evaluate the performance of the factorization ansatz as a function of radiation type for the processes $\Pd \Pg \to \PZ \Pd \Pg \Pg \Pg$ and $\Pd \Pdbar \to \PZ \Pg \Pg \Pg \Pg$, including both initial- and final-state radiation. We find that final-state gluon radiation exhibits higher approximation accuracy in both cases. This trend can be attributed to combinatorial effects associated with the number of gluons, which increases the probability of identifying a softer or more collinear pair.

%We do explain this by the number of gluon present in the final state, leading to softer and/or more collinear event within the splitting kernel.

%%%%%%%%%%%%%%%%%%%%%%%%%%%%%%%%%%%%%%%%%%%%%%%%%%%%
\subsection{Double factorization}

Using the CS formalism, we can iteratively absorb jets from the final state, thereby reducing the number of particles in the final configuration by one at each step. This procedure allows us to transition from a final state with $n$-particles to one with $(n-2)$-particles through successive reductions.

In the first step, one radiation is absorbed from the $n$-particle final state, resulting in a reduced process with $(n-1)$-particles. Then, a second radiation is absorbed to arrive at a $(n-2)$-particle state. The second reduction involves a splitting that depends on the reduced particle $\tilde{p}_{ij}$ produced during the first absorption. The approximated amplitude can thus be expressed as
\begin{align}
A_{n} & \approx A_{n-1} \cdot F_{ij,k}^r \approx A_{n-2} \cdot F_{\widetilde{ij} l,k^\prime}^{r} \cdot F_{ij,k}^r\;,
\end{align}
where $F_{\widetilde{ij} l,k^\prime}^{r}$ is the splitting kernel corresponding to the second reduction, and it depends on the momenta of the particles $\tilde{p}_{ij}$, $p_l$, and $p_{k^\prime}$. The kinematics of the second reduction are therefore influenced by the first one, since one of the particles $p_l$ or $p_{k^\prime}$ may also serve as the spectator in the initial reduction.

In principle, one could select a radiation for the second reduction that does not involve the reduced particle $\tilde{p}_{ij}$ generated in the first step. However, in our study, we observed better performance from networks in which both reductions share the same reduced particle. For this reason, we restrict the radiation combinations to those that involve $\tilde{p}_{ij}$ in both steps. Similarly, although it is possible to mix different radiation types between the first and second reduction, we did not observe any significant benefit from doing so. Therefore, for simplicity, we limit our analysis to double factorizations constructed from identical radiation types.

%----------------------------------------
\begin{figure}[tb!]
    \includegraphics[width=0.49\linewidth]{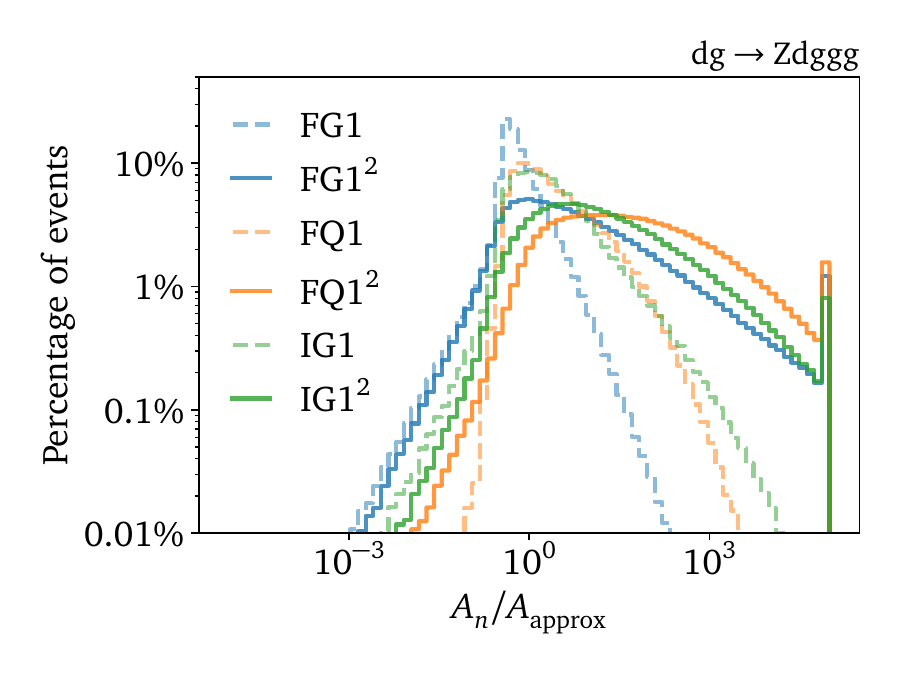}
    \includegraphics[width=0.49\linewidth]{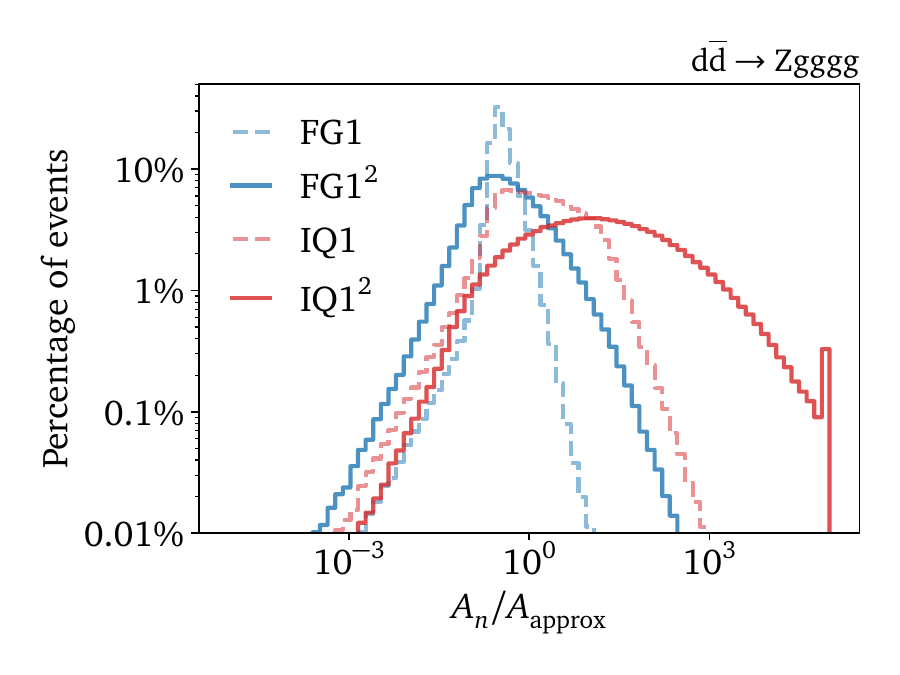}
    \caption{Approximation quality comparison between single and double factorizations for $\Pd \Pg \to \PZ \Pd \Pg \Pg \Pg$ (left) and for $\Pd \Pdbar \to \PZ \Pg \Pg \Pg \Pg$ (right).}
    \label{fig:Plot_Fact_Approx_double}
\end{figure}
%----------------------------------------

In Fig.~\ref{fig:Plot_Fact_Approx_double}, we compare the approximation accuracy of the double factorization with that of the single factorization. To distinguish the notation, we denote the double factorization with a superscript square, for example FG1$^2$. Overall, the double approximation tends to exhibit reduced accuracy. While the central value of this approximation is of limited relevance -- since it can be corrected by the neural network -- it is worth noting that the double factorization estimate shows a noticeably broader spread. This implies that the neural network will need to handle a wider range of values, which may lead to a decrease in performance. Nevertheless, evaluating the double approximation is approximately ten times faster (see Tab.~\ref{tab:MG_speed}), which could make it useful for specific applications (see Section~\ref{sec:speedup}).

%----------------------------------------
\begin{table}[b!]
    \centering
    \setlength{\tabcolsep}{8pt}
    \begin{tabular}{c c} 
        \toprule
        Feature & Evaluation time [s] \\ 
        \midrule
        $A_n$ & $1.3 \cdot 10^{-3}$ \\ 
        $A_{n-1}$ & $1.0 \cdot 10^{-4}$ \\ 
        $A_{n-2}$ & $1.2 \cdot 10^{-5}$ \\ 
        $c_{\theta}$ & $2.0 \cdot 10^{-8}$ \\
        \bottomrule
    \end{tabular}
    \caption{Summary of the average evaluation time per event for the process $\Pd \Pdbar \to \PZ \Pg \Pg \Pg \Pg$ and the corresponding reduced processes. }
    \label{tab:MG_speed}
\end{table}
%----------------------------------------

%\clearpage 
%%%%%%%%%%%%%%%%%%%%%%%%%%%%%%%%%%%%%%%%%%%%%%%%%%%%
\section{Machine-learning surrogate}
\label{sec:ml}

Our aim is to develop of a fast and accurate neural surrogate that leverages the factorized structure of the amplitude to approximate the full result. The factorization ansatz introduced in the previous section isolates the dominant singular behavior, leaving the network to learn a smooth correction factor with reduced variance and without large variations associated with on-shell resonances or explicit soft/collinear singularities.

In this section, we outline our machine-learning setup. We begin in Subsection~\ref{sec:loss} by discussing the heteroscedastic loss and its relation to our performance metric. In Subsection~\ref{sec:architecture}, we describe the architecture employed for the various networks throughout this work and explain how amplitude predictions from different networks can be combined.

%%%%%%%%%%%%%%%%%%%%%%%%%%%%%%%%%%%%%%%%%%%%%%%%%%%%
\subsection{Preprocessing and heteroscedastic loss}
\label{sec:loss}

Our training targets are exact theoretical amplitudes $A_\text{true}(x)$ for given phase-space points $x$, as defined in Eq.\eqref{eq:def:amplitude}.  
Rather than regressing directly on the amplitude itself, which can span many orders of magnitude across phase space, we train the network on a logarithmic representation,
\begin{align}
\label{eq:standardization}
\begin{split}
\ell_\text{true}(x) 
&= \frac{\log A_\text{true}(x) - \mu_{\text{train}}}{s_{\text{train}}} \\
\mwith \qquad
\mu_{\text{train}} &= \Langle\log A_\text{true}(x)\Rangle_{x\sim D_\text{train}} \qquad
s^2_{\text{train}} = \Langle\left(\log A_\text{true}(x)-\mu_{\text{train}}\right)^2\Rangle_{x\sim D_\text{train}}\;,
\end{split}
\end{align}
including a linear standardization computed on the training data, where $\langle \cdot \rangle_{x \sim D_{\text{train}}}$ denotes an average over the training set. All network predictions are made in this standardized log-amplitude space and are mapped back to amplitude level only when evaluating performance or deploying the surrogate.
This preprocessing stabilizes training and renders the learning problem numerically well-conditioned.

In this setup, there is no intrinsic measurement noise or stochasticity in the training data. Deviations between predictions and targets therefore arise purely from epistemic sources, reflecting our incomplete knowledge of the true amplitude function. These include (i) \emph{finite-data and training-related effects} due to insufficient coverage of phase space and variability in the optimization procedure, which vanish in the limit of infinite data and perfect training, as well as (ii) a residual \emph{model bias} originating from limitations of the factorization ansatz or the expressiveness of the neural network, which does not vanish even in the infinite-statistics limit.

We nevertheless employ a heteroscedastic-style objective to let the network predict, alongside its mean log-amplitude prediction $\ell_\theta(x)$, an input-dependent reliability proxy $\sigma_\theta(x)$, where $\theta$ denotes the trainable network weights. 
The loss can then be written as
\begin{align}
    \loss_\text{het}
    = \left\langle \frac{\left(\ell_\text{true}(x) - \ell_\theta(x)\right)^2}
                     {2\,\sigma^2_\theta(x)}
      + \log \sigma_\theta(x) \right\rangle_{x\sim D_\text{train}}\;.
      \label{eq:het_loss}
\end{align}
Compared to a standard mean-squared-error loss, this objective allows the network to dynamically downweight difficult regions of phase space during training. 
While Eq.\eqref{eq:het_loss} matches the algebraic form of a Gaussian likelihood, $\sigma_\theta$ is not interpreted as data (aleatoric) noise. Instead, it serves as a learned weighting term during optimization, with the following practical benefits:
\begin{enumerate}
    \item Downweighting of rare or hard-to-learn configurations, leading to improved global accuracy and training stability.
    \item Per-event confidence proxies that can be exploited in uncertainty-weighted combinations of multiple models, yielding more robust ensemble predictions.
\end{enumerate}
Importantly, $\sigma_\theta(x)$ is \emph{not} a measure of epistemic uncertainty in the statistical sense. 
Since it is learned jointly with the mean prediction $\ell_\theta(x)$, it absorbs a mixture of 
finite-data effects, optimization variability, and residual model bias.

More fundamentally, attempting to represent epistemic uncertainty by a point-wise quantity such as $\sigma_\theta(x)$ would destroy correlations between different phase-space points. Such correlations are essential for the consistent propagation of uncertainties to
derived quantities, for example integrated or differential cross sections. To retain them, epistemic uncertainty must instead be extracted from \emph{between-model} variability, for instance by using independently trained ensembles or Bayesian neural networks (BNNs)~\cite{bnn_early, bnn_early2,bnn_early3,Bollweg:2019skg,Kasieczka:2020vlh,ATLAS:2024rpl}, which approximate sampling from the posterior $p(\theta | D_{\text{train}})$ and thus yield coherent, correlated uncertainty estimates across phase space.

We also tested BNNs, which are, in principle, well suited for estimating epistemic uncertainties. In practice, however, they produce central predictions comparable to those of our standard networks, while introducing increased training instability, higher model complexity, and slower inference. Moreover, when training a BNN with the heteroscedastic loss, the $\sigma_\theta(x)$ term tends to absorb most epistemic variation, consistent
with the expectation that these contributions become entangled in the zero-noise limit. A proper disentangling of epistemic components would require a more structured training procedure, such as the multi-step approach outlined in Ref.~\cite{yi2025cooperativebayesian}.
Since the focus of this work is on demonstrating a proof of concept for factorization-based amplitude surrogates, we leave such developments to future work.

\subsubsection*{Performance measure}

For evaluation and physics validation, we quantify surrogate accuracy at the amplitude level through the relative deviation
\begin{align}
    \Delta(x) = \frac{A_\text{true}(x) - A_\theta(x)}{A_\text{true}(x)}\;,
    \label{eq:delta_def}
\end{align}
which will be the primary performance metric used in Sec.~\ref{sec:results}.
For small deviations, training in log-amplitude space is directly related to this quantity, since
\begin{align}
    \loss_\text{het}\propto (\ell_\theta(x)-\ell_\text{true}(x))^2
    = \frac{1}{s^2_\text{train}}
      \log^2\!\left(1-\Delta(x)\right)
    \simeq \frac{\Delta^2(x)}{s^2_\text{train}}
    + \mathcal{O}(\Delta^3)\,.
\end{align}
This means minimizing the mean-squared error in log space corresponds approximately to minimizing the squared relative deviation $\Delta^2$, while avoiding numerical instabilities associated with large dynamic ranges in $A(x)$.
The heteroscedastic loss therefore acts as a stabilized and uncertainty-weighted version of relative-error minimization.

%%%%%%%%%%%%%%%%%%%%%%%%%%%%%%%%%%%%%%%%%%%%%%%%%%%%
\subsection{Factorization network}
\label{sec:architecture}

%----------------------------------------
\begin{figure}[b!]
    \centering
    \includegraphics[width=0.7\linewidth]{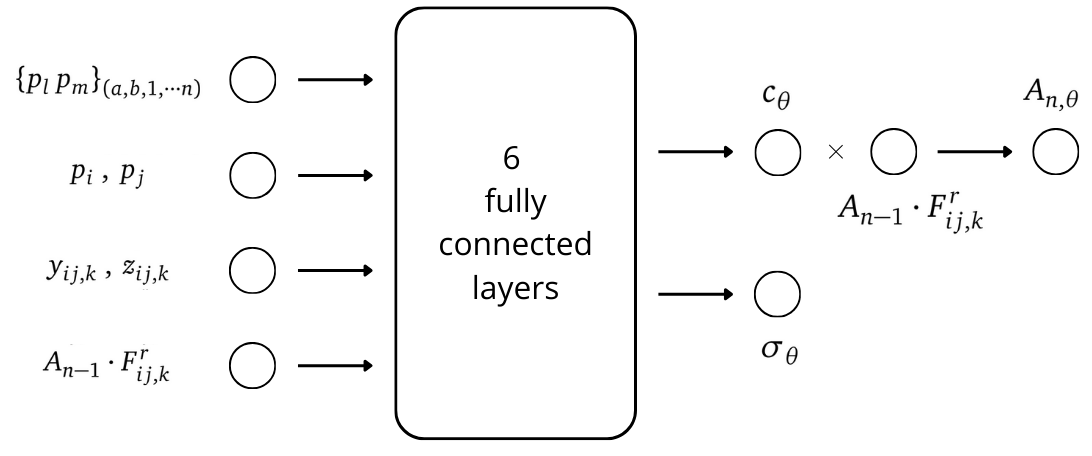}
    \caption{Representation of the single factorization neural network.}
    \label{fig:Sketch_NN}
\end{figure}
%----------------------------------------

Each factorization network uses a factorization approximation to reconstruct the full amplitude starting from a reduced one. Specifically, the network outputs a correction factor $c_{\theta}$, which rescales the factorized approximation to match the true value. For a single factorization model, the predicted amplitude takes the form:
\begin{align}
A_{n,\theta} = c_{\theta} \cdot A_{n-1} \cdot F_{ij,k}^r
\label{eq:FaCS_ansatz_single}
\end{align}
where the radiation label $r$ specifies the radiation type. 
For a double factorization model, the predicted amplitude is:
\begin{align}
A_{n,\theta} = c_{\theta} \cdot A_{n-2} \cdot F_{\widetilde{ij} l,k^\prime}^{r} \cdot F_{ij,k}^r
\label{eq:FaCS_ansatz_double}
\end{align}
As a baseline, we also consider a \emph{No Factorization} (NoFa) model, which predicts the amplitude directly without using any analytical factorization structure. In this case, 
\begin{align}
    A_{n,\theta} = c_{\theta}
\end{align}
where $c_{\theta}$ is obtained by de-processing the raw network output $c_{\theta}^\prime$ using the same transformation applied to the training target amplitudes $c_{\theta} = \exp{ \left( c_{\theta}^\prime \cdot s_{\text{train}} + \mu_{\text{train}} \right)}$, allowing the network to effectively cover several orders of magnitude. 

In our study, we also explored a multi-radiation factorization model, in which a single network predicts the full amplitude as a linear combination of single-radiation approximations. The ansatz for this model is
\begin{align}
\label{eq:multiradiation}
    A_{n,\theta} = \sum_{r}^{N_r} c^{r}_{\theta} \cdot A^r_{n} \cdot F_{ij,k}^r
\end{align}
where the sum over $r$ includes the $N_r$ selected radiations. 
The reduced amplitude $A_{n}^r$ differs for each radiation and must be computed $N_r$ times for each event. In our tests, this approach did not show a clear improvement in accuracy compared to the single-radiation model. Given its higher computational cost, we chose not to include this model in the final results.
 
%----------------------------------------
\begin{table}[tb!]
    \centering
    \setlength{\tabcolsep}{10pt}
    \begin{tabular}{l c} 
        \toprule
        NN hyper-parameters & Value  \\ 
        \midrule
        Nodes per layer  & [128, 256, 256, 128, 128, 64] \\
        Total model parameters & $1.6 \cdot 10^{5}$ \\
        Activation function & Gelu \\ 
        Optimizer & AdamW \\
        Initial learning rate & $10^{-3}$ \\
        Final learning rate & $10^{-8}$ \\
        Learning rate scheduler & CosineAnnealing \\
        Max epochs & $1000$ \\
        Batch size & $256$ \\
        Training events & $5 \cdot 10^{5}$ \\
        Validation events & $1 \cdot 10^{5}$ \\
        \bottomrule
    \end{tabular}
    \caption{Summary of the hyper-parameters used in a factorization neural network.}
    \label{tab:NN_params}
\end{table}
%----------------------------------------

%----------------------------------------
\begin{table}[b!]
    \centering
    \setlength{\tabcolsep}{10pt}
    \begin{tabular}{l c} 
        \toprule
        Input feature & Variable  \\ 
        \midrule
        Kinematic log-invariants & $\{\log p_l p_m\}_{l,m}$ \\ 
        Emitter and emitted four-momenta & $p_{i}, p_j$ \\ 
        Radiation variables & $y_{ij, k}, z_{ij, k}$ \\ 
        Log-factorization factors & $\log \left(A_{n-1}\, F_{ij,k}^r\right)$ \\
        \bottomrule
    \end{tabular}
    \caption{Summary of input features used in a single factorization model.}
    \label{tab:NN_input}
\end{table}
%----------------------------------------

The technical details of our neural-network architecture are summarized in two tables. Table~\ref{tab:NN_params} lists the hyperparameters used for all networks in this work, while Table~\ref{tab:NN_input} specifies the input features for a single-factorization network. The kinematic log-invariants and the logarithmic factorization inputs are also standardised, in order to account for their disparate orders of magnitude.
To guide the model toward the most relevant configurations, we reorder the momenta of all possible emitter and emitted particles for the corresponding radiation type before computing the kinematic invariants. This rearrangement is based on the singularity associated with each radiation and encodes the radiation rank into the particle ordering.

For the NoFa model, no reordering is applied, and no factorization information is provided as input. In addition, for the double-factorization model, two additional outputs are included to provide to the network the extra radiation variables.

%%%%%%%%%%%%%%%%%%%%%%%%%%%%%%%%%%%%%%%%%%%%%%%%%%%%
\subsubsection*{Network ensembling}

Ensemble neural network models combine the predictions of multiple individual networks to improve robustness and accuracy. By aggregating diverse models, ensemble methods reduce over-fitting and provide a broader, more reliable representation of the underlying physical processes. In this work, we employ an ensemble of simple neural networks, each based on a distinct factorization approximation. 
It differs from the multi-radiation single network (Eq.\ref{eq:multiradiation}), where the same network predicts all the correction factors of the ansatz.
%, and the prediction does not depend on the predicted uncertainties of the radiations. 

We define our ensemble model as a linear combination of factorization models.
Given a set of factorization models $\{ 1, \dots, n_{\text{mod}} \}$, a model $j$ predicts for a given event $i$ an output $(A_{\theta_j}^{i}, \sigma_{\theta_j}^{i})$, where $A_{\theta_j}^{i}$ is the predicted amplitude for the event $i$ and $\sigma_{\theta_j}^{i}$ is the corresponding predicted uncertainty. 
The predicted ensemble amplitude is given by
\begin{align}
    A_{\theta_e}^{i} = \sum_{j=1}^{n_{\text{mod}}} w_j^{i} \cdot A_{\theta_j}^{i} \;,
    \label{eq:FaCS_pred_ens}
\end{align}
where the weights $w_j^i$ are computed from the inverse of the covariance matrix $C_i^{-1}$ of the model predictions~\cite{Lyons:1988rp,Valassi:2003mu}
\begin{align}
    w_j^i = \frac{\sum_k (C_i^{-1})_{jk}}{\sum_{j,k} (C_i^{-1})_{jk}} \;.
\end{align}
To estimate $C_i$, we first determine the correlation matrix $\rho$ between two different models. The correlation $\rho_{ab}$ is defined as the average correlation between the per-point accuracies of models $a$ and $b$ over the entire validation dataset. This correlation is treated as a global quantity and kept fixed across phase space. We also investigated a per-point, phase-space–dependent correlation, but found neither improvement nor degradation in performance. Defining $\rho_{ab}$ in terms of accuracy allows us to quantify whether two models tend to make similar prediction errors.
The covariance between models $a$ and $b$ for event $i$ is then given by
\begin{align}
    C_{i_{ab}} = \rho_{ab} \cdot \sigma_{\theta_b}^i \cdot \sigma_{\theta_b}^i 
\end{align}
from which we obtain the inverse matrix $C_i^{-1}$.
Finally, the predicted uncertainty for the ensemble model is
\begin{align}
    \sigma_{\theta_e}^i = \frac{1}{\sqrt{\sum_{j,k} (C_i^{-1})_{jk}}} \;.
\end{align}

\clearpage 
%%%%%%%%%%%%%%%%%%%%%%%%%%%%%%%%%%%%%%%%%%%%%%%%%%%%
\section{Application to Z with multi-jet production}
\label{sec:results}

As previously noted, we benchmark our CS-based ansatz using Z+jet production. This choice is mainly motivated by its use in previous surrogate studies~\cite{Maitre:2021uaa,Brehmer:2024yqw}, which facilitates direct comparison with existing approaches. Nonetheless, our method will exhibit even greater advantages for Beyond Standard Model processes accompanied by jets, or more generally for scenarios involving multiple resonances and/or scales in conjunction with jets. Within the Standard Model, particularly suitable candidates include vector-boson fusion with jets or fully decayed top-quark production with jets. A comprehensive treatment of these processes, together with full automation of the proposed methodology, is deferred to future work.

\subsection{Performance for quark-gluon initial state}

We begin by comparing single factorization models applied to different types and ranks of radiation within the same process. Specifically, we consider the process
$\Pd \Pg \to \PZ \Pd \Pg \Pg \Pg$, and evaluate the following radiation configurations: FG1, FQ1, IG1, IQ1, for the radiation types, and FG1, FG2, FG3, for the radiation ranks.

\begin{figure}[b!]
    \includegraphics[width=0.49\linewidth]{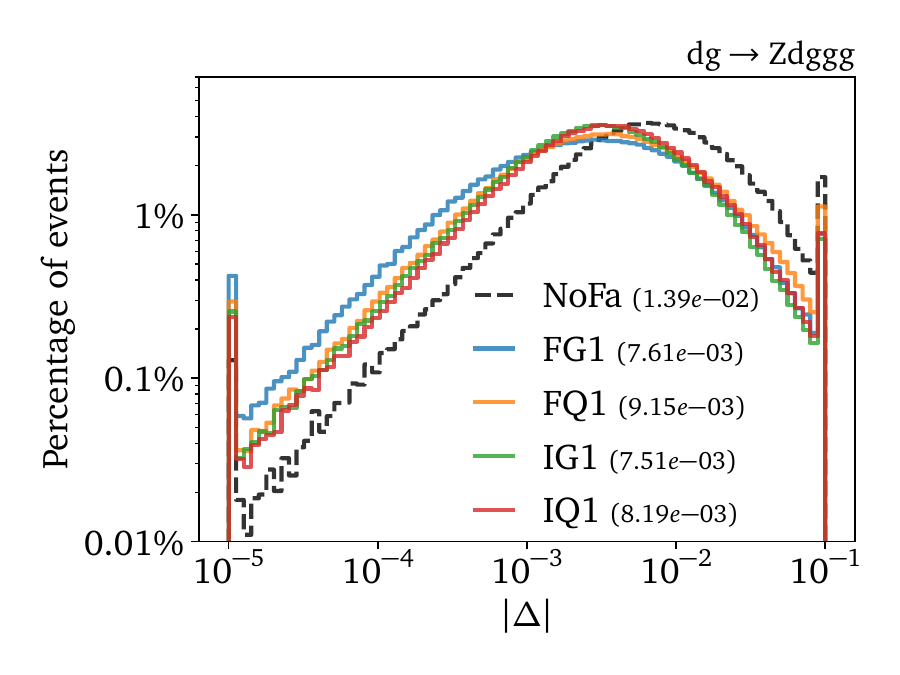}
    \includegraphics[width=0.49\linewidth]{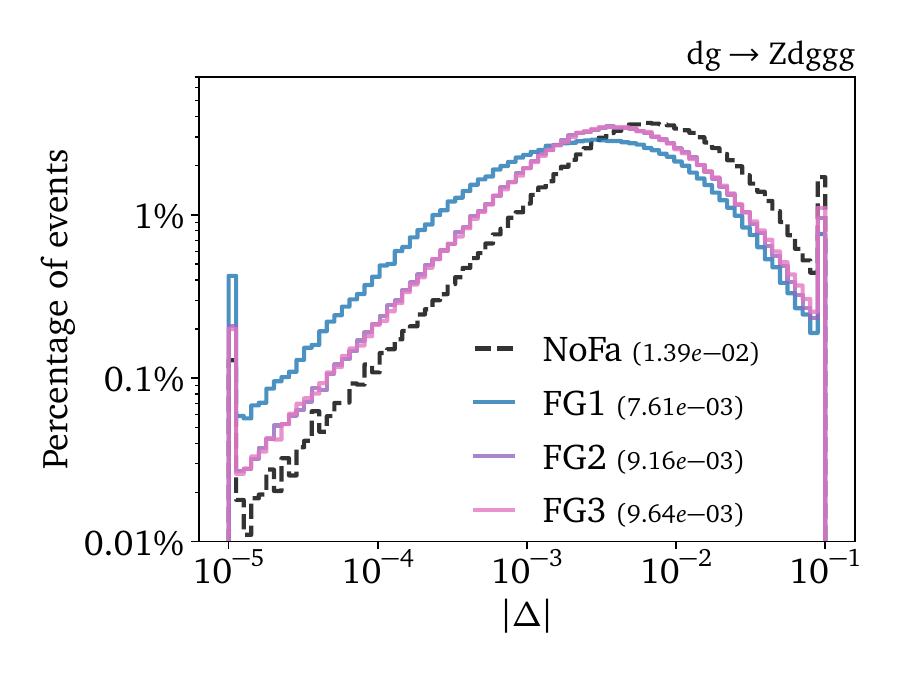}
    \caption{Single factorization model accuracy comparison between different radiation types (left), and different radiation ranks (right) for the process $\Pd \Pg \to \PZ \Pd \Pg \Pg \Pg$. The values in parentheses are the mean accuracies over the whole test dataset.} 
    \label{fig:Plot_NN_Acc_dg_zd3g_single}
\end{figure}

Figure~\ref{fig:Plot_NN_Acc_dg_zd3g_single} compares the performance of the models introduced above. First, we observe that all radiation-based models outperform the NoFa model, confirming the benefit of incorporating the factorization structure.
In the left panel, the factorization models achieve similar average accuracy despite originating from different underlying factorization qualities. The two bests models are here FG1 and IG1 with very similar mean accuracy but quite different shape: IG1 being more peak (i.e. consistent with the average) while FG1 has a broader shape, with especially more event with better precision than $10^{-3}$. Both models have quite similar rate of events with low accuracy ($>10^{-2}$).   

The picture is simpler in the right panel, where we present the accuracy as a function of the approximation rank. Selecting the most singular approximation yields a clear improvement in accuracy, both in terms of the mean and the overall distribution. In contrast, there is no noticeable difference between choosing the second or third rank, indicating that the critical information lies primarily in the first-rank emission. This provides an a posteriori indication that using only one reduction step—specifically the most singular one—is a reasonable choice.

%----------------------------------------
%----------------------------------------

%-------------------------------
\begin{figure}[t!]
\centering
    \includegraphics[width=0.49\linewidth]{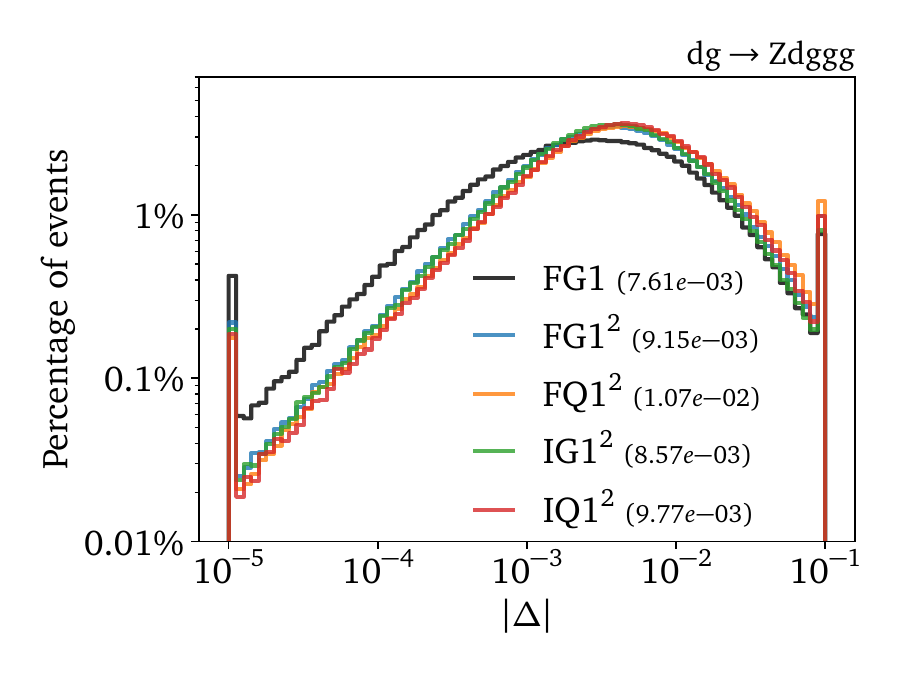}
    \caption{Double factorization model accuracy comparison between different radiation types for the process $\Pd \Pg \to \PZ \Pd \Pg \Pg \Pg$. The values in parentheses are the mean accuracies over the whole test dataset.} 
    \label{fig:Plot_NN_Acc_dg_zd3g_double}
\end{figure}
%-------------------------------

We now compare the performance of the double factorization models across different radiation types: FG1$^2$, FQ1$^2$, IG1$^2$ and IQ1$^2$.
In Fig.~\ref{fig:Plot_NN_Acc_dg_zd3g_double}, we report the corresponding accuracy distributions. The double factorization models exhibits reduced accuracy compared to their single-factorization counterparts. This was obviously expected since the underlying approximation was much less precise to start with. Additionally, one can also see a much weaker dependence in the kernel used, which is also not very surprising given the distribution to learn by the neural network (Fig.~\ref{fig:Plot_Fact_Approx_double}) which shows that all curves are indeed quite similar.  
It is noteworthy that, for FG radiation, the double-factorization model FG1$^2$ exhibits performance that is very close to that of the single-factorization model of second rank, FG2. Finally, when comparing the double-factorization model to the NoFa baseline for this process, the improvement remains modest.

\subsubsection*{Ensemble factorization models}
After evaluating the performance of individual radiation models, we now investigate how their predictions can be combined through network ensembles to improve accuracy and robustness. As previously discussed, each radiation model captures different aspects of the same underlying process, and even models with similar accuracy may encode different information due to the type and rank of the radiation. 
While this motivates the use of ensemble methods to leverage the diversity among models, a large part of that information is redundant and one need to be careful about the correlations between the various predictions.

\begin{figure}[b!]
    \includegraphics[width=0.49\linewidth]{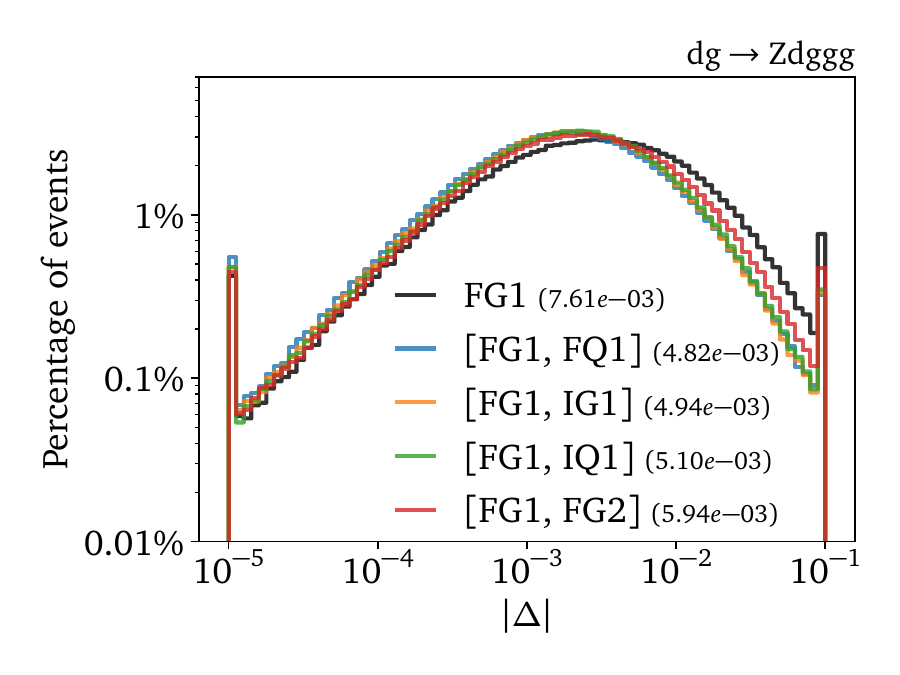}
    \includegraphics[width=0.49\linewidth]{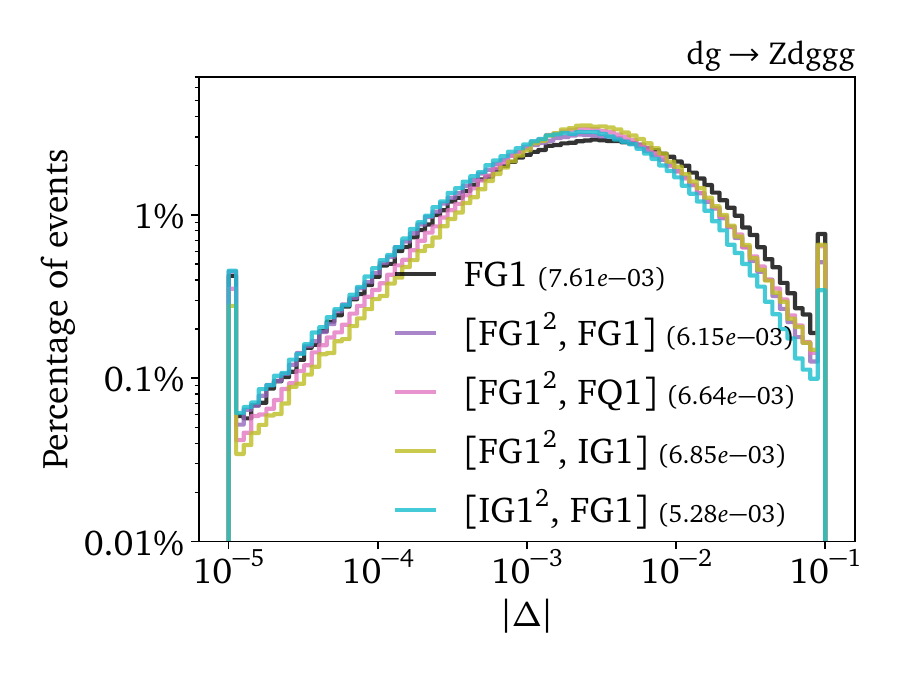}
    \caption{Ensemble factorization model accuracy comparison between single factorization (left), and double and single factorization (right) for the process $\Pd \Pg \to \PZ \Pd \Pg \Pg \Pg$. The values in parentheses are the mean accuracies over the whole test dataset.} 
    \label{fig:Plot_NN_Acc_dg_zd3g_ensemble}
\end{figure}

As described before, we define an network ensemble as a combination of independently trained networks. In this context, how the individual predictions are recombined is crucial. A naive recombination that is independent of the predicted uncertainty can lead to suboptimal results. In our case, we observed that the uncertainty-weighted recombination described in Eq.\eqref{eq:FaCS_pred_ens} consistently outperforms any of the individual single models. From Fig.~\ref{fig:Plot_NN_Acc_dg_zd3g_ensemble}, we confirm that each ensemble achieves a better mean accuracy than its constituent individual models. 

Interestingly, the overall accuracy of the ensemble does not depend solely on the accuracies of the individual models, but also on the radiation types and ranks used to construct the ensemble.  
Combining the most accurate individual models does not always yield the best ensemble performance, as shown by the ensemble [FG1, FQ1] achieving a better accuracy than [FG1, IG1], despite IG1 having better performance. 
This behavior highlights that networks trained on different radiations tend to extract complementary information from the same dataset, and that low error correlation, or model diversity, is a key factor for achieving strong ensemble results. We also experimented with modifying the loss and the learning algorithm to encourage such complementarity and improve accuracy, but without significant success.

%-------------------------------
%-------------------------------

On the right panel, we compare ensemble models of double factorization combined with single factorization one. These ensembles, even if yielding a lower accuracy due to the approximation used, will actually have better speed-up factor due to faster evaluation time.

Similarly, we observe that ensembles composed of three models outperform those composed of only two, although the performance gain is less significant than the improvement obtained when moving from a single model to an ensemble of two. 

\subsection{Performance for quark-quark initial state}

So far, we have evaluated the performance of single and ensemble models on the process $\Pd \Pg \to \PZ \Pd \Pg \Pg \Pg$ to investigate how the accuracy depends on the chosen factorization model. In this section, we extend the study to a different process: $\Pd \Pdbar \to \PZ \Pg \Pg \Pg \Pg$ with the main difference compare to the previous section is that all the gluon are in the final state.

\subsubsection*{Single and double factorization models}

\begin{figure}[htp]
    \includegraphics[width=0.49\textwidth]{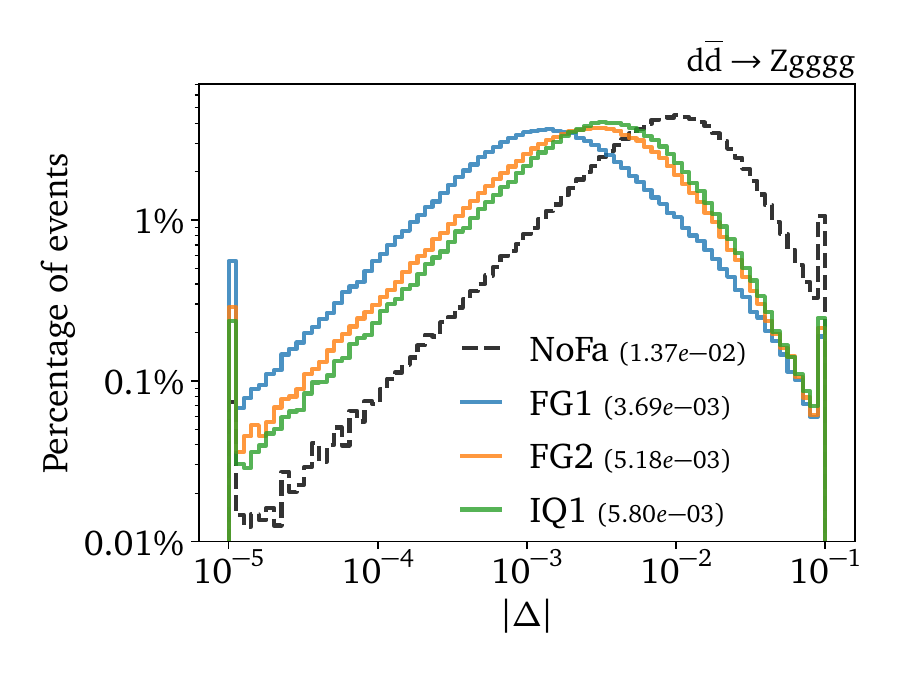}
    \includegraphics[width=0.49\textwidth]{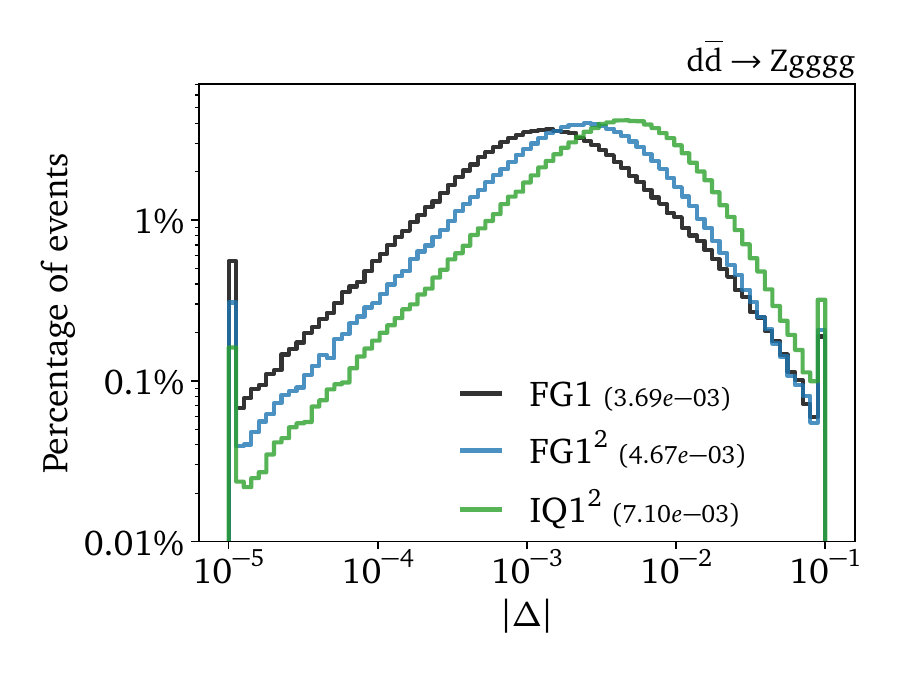}
    \caption{Accuracies (left) and cumulative accuracy (right) of single radiation models on $\Pd \Pdbar \to \PZ \Pg \Pg \Pg \Pg$ process. On the left, the values in parentheses are the mean of the distribution.}
    \label{fig:Plot_NN_Acc_ddx_z4g_indiv}
\end{figure}

We begin by examining the performance of the individual factorization models. The left panel of Fig.~\ref{fig:Plot_NN_Acc_ddx_z4g_indiv} shows the single-factorization results. Comparing this plot with the corresponding one for the previous process (Fig.~\ref{fig:Plot_NN_Acc_dg_zd3g_single}), we observe that the NoFa model achieves nearly identical accuracy in both cases. This is not the case for the factorization models: in particular, the best-performing model, FG1, attains substantially higher accuracy. This improvement appears to result from having all gluons—which dominate the radiation pattern—in the final state, thereby enhancing the effectiveness of the factorization ansatz.

The right panel of Fig.~\ref{fig:Plot_NN_Acc_ddx_z4g_indiv} shows the accuracy of the double-factorization model. Here as well, the performance remains significantly better than the NoFa baseline. Moreover, similarly to the $\Pd \Pg \to \PZ \Pd \Pg \Pg \Pg$ case, we observe that the accuracy achieved by FG1$^2$ is nearly identical to that of the single-factorization FG2 model.

\subsubsection*{Ensemble factorization models}
\begin{figure}[b!]
    \includegraphics[width=0.49\textwidth]{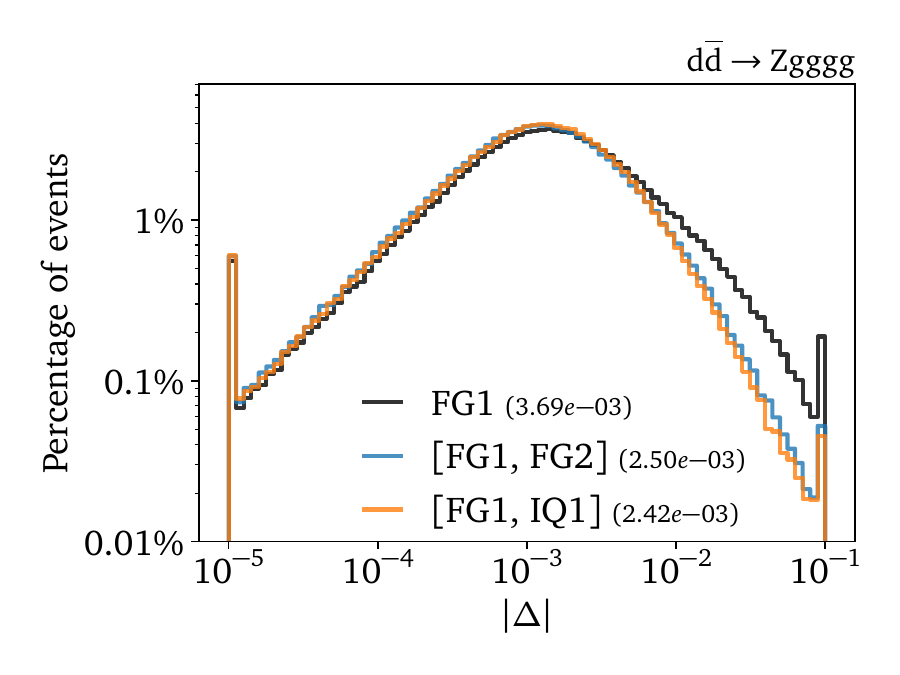}
    \includegraphics[width=0.49\textwidth]{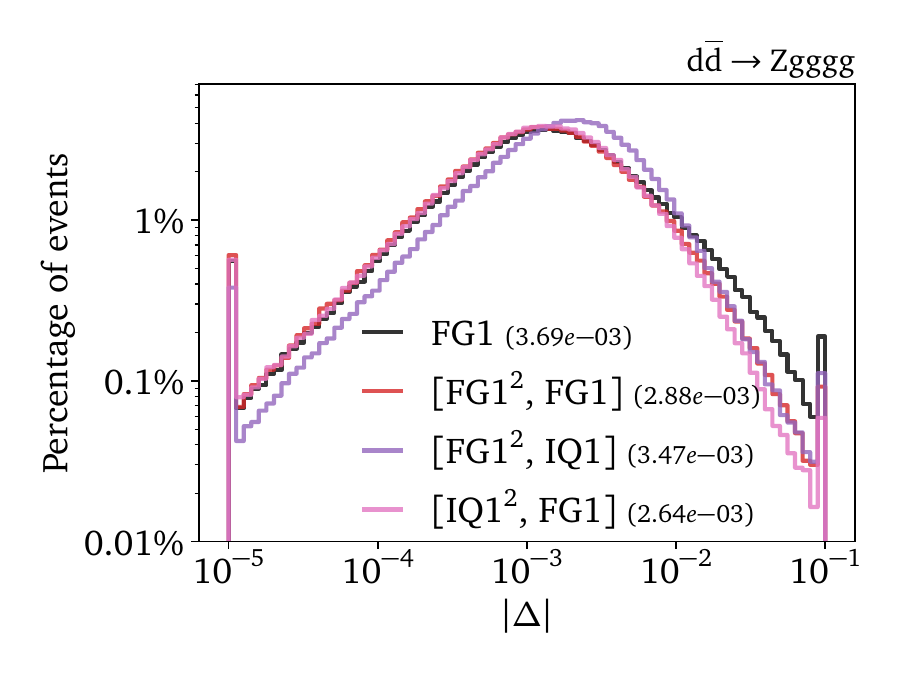}
    \caption{Ensemble factorization model accuracy comparison between single factorization (left), and double and single factorization (right) for the process $\Pd \Pdbar \to \PZ \Pg \Pg \Pg \Pg$. The values in parentheses are the mean accuracies over the whole test dataset.}
    \label{fig:Plot_NN_Acc_ddx_z4g_ensemble}
\end{figure}

Finally, we examine the accuracy of the ensemble models for the process $\Pd \Pdbar \to \PZ \Pg \Pg \Pg \Pg$. As in the previous process, we confirm that the ensemble models outperform the individual ones across the entire test dataset. In the left panel of Fig.~\ref{fig:Plot_NN_Acc_ddx_z4g_ensemble}, we directly compare each ensemble with its best corresponding individual model. We observe that the ensembles reduce the large-error tail more significantly than the small-error tail, indicating that combining different models is especially beneficial for events that are otherwise difficult to predict accurately.

%-------------------------------
%-------------------------------

The right panel of Fig.~\ref{fig:Plot_NN_Acc_ddx_z4g_ensemble} shows ensembles that combine double- and single-factorization models. In this case, the best-performing ensemble is [IG1$^2$, FG1]. This result suggests that the most effective strategy is not to combine the two most accurate models, but rather those that provide complementary information. Furthermore, this ensemble achieves an accuracy comparable to the best ensembles composed exclusively of single-factorization models, while offering faster evaluation times.

This process corresponds to the one used in \cite{Brehmer:2024yqw}, which employs a Lorentz-equivariant network. A direct comparison of the achieved accuracy is therefore possible. In that work, accuracy is not reported directly; instead, the authors use the mean squared error (MSE) on the logarithmic standardized amplitude as a measure of network performance. For comparison purposes, we report in Table~\ref{tab:mse_comp} the corresponding values for several of our models. These results indicate that the achieved accuracy is in the same ballpark as that of the L-Gatr network, which remains more performant under this metric than any of our single-factorization models. However, our most accurate ensemble network surpasses it.

%----------------------------------------
\begin{table}[tb!]
    \centering
    \setlength{\tabcolsep}{10pt}
    \begin{tabular}{l c c} 
        \toprule
        Neural network model & Mse & mean$(|\Delta|)$ \\ 
        \midrule
        FG1  & $1.5 \cdot 10^{-5}$ & $3.69 \cdot 10^{-3}$ \\
        FG1$^2$ & $ 1.8\cdot 10^{-5}$ & $4.67 \cdot 10^{-3}$ \\
        $[$FG1, FG1$^2]$ & $8.7 \cdot 10^{-6}$ & $2.88 \cdot 10^{-3}$ \\
        L-GATr*  & $ \sim8.4 \cdot 10^{-6}$ & \\
        $[$FG1, FG2$]$ & $5.9 \cdot 10^{-6}$ & $2.50 \cdot 10^{-3}$ \\
        $[$FG1, IQ1$]$ & $5.5 \cdot 10^{-6}$ & $2.42 \cdot 10^{-3}$ \\

        \bottomrule
    \end{tabular}
    \caption{Comparison of the various network models in terms of logarithmically standardized mean squared error (mse), including L-GATr \cite{Brehmer:2024yqw}(trained on $4 \cdot 10^5$ samples). The last column reports, for our models, the mean accuracy metric used in this paper. Model are ordered from less precise to the most accurate one (according to the mse).
    \label{tab:mse_comp}}
\end{table}
%----------------------------------------

%\clearpage
%%%%%%%%%%%%%%%%%%%%%%%%%%%%%%%%%%%%%%%%%%%%%%%%%%%%
\section{Achievable speed-up without double-unweighting}
\label{sec:speedup}

Having a fast and accurate surrogate is only useful if we can use is it practically. While we here focus on how to make LO phase-space integration faster, the algorithm that we will describe in this section can generally be applied to any computation relying on evaluating amplitudes. In the literature~\cite{Danziger:2021eeg,Herrmann:2025nnz, Villadamigo:2025our} a common approach when using surrogates is to use it within importance sampling, while relying on standard hit and miss to generate event following the surrogate density function, which is conceptually equivalent to other importance sampling algorithm (based on ML or not)~\cite{Lepage:1977sw, Lepage:1980dq,Lepage:2020tgj, Heimel:2022wyj, Mattelaer:2021xdr,Heimel:2024wph,Maltoni:2002qb,vanHameren:2007pt,Kleiss:1985gy, Kleiss:1994qy, Klimek:2018mza,Gao:2020vdv,Heimel:2023ngj,Janssen:2025zke,Bothmann:2020ywa,Bothmann:2023siu, Bothmann:2025lwg}. In those methods, even in the perfect scenario, one need to evaluate at least one amplitude (and typically much more) for each unweighted event produced.

In this paper, we propose using the surrogate directly in place of the full computation when its estimated error is sufficiently small. This approach, described in more detail below, can significantly reduce the number of evaluations of the full amplitude—potentially eliminating them entirely at the generation stage. The main drawback is that it introduces an additional source of numerical uncertainty associated with the precision of the surrogate itself.

An often-cited concern with such a strategy is that the tails of distributions might be severely misrepresented. In practice, however, most observables $\mathcal{O}$ correspond to cross sections integrated over restricted regions of phase space. For a binned (differential) observable defined by a phase-space region $\Omega\subset\Phi$, for instance corresponding to a bin in a low-$p_T$ tail, we can write
\begin{align}
  \mathcal{O}
  \equiv
  \int_{\Omega} \d x\,\frac{\d\sigma}{\d x}
  \approx
  \frac{1}{N}\sum_{i\in\Omega} w_i \, ,
\end{align}
where the sum runs over all Monte-Carlo events that fall within the phase-space region $\Omega$, and the weights $w_i$ denote the usual event
weights, including the amplitude, the parton distribution functions, and the Jacobian associated with the phase-space measure.
In the ideal case, the errors on the individual weights $w_i$ are fully uncorrelated and thus
\begin{align}
   (\Delta \mathcal{O})^2 &=\frac{1}{N^2}\sum_i (\Delta w_i)^2\;.
\end{align}
In this case, the relative uncertainty on any observable $\mathcal{O}$ vanishes in the infinite–statistics limit
\begin{align}
\frac{(\Delta \mathcal{O})^2}{\mathcal{O}^2} &=\frac{\sum_i (\Delta w_i)^2}{(\sum_i w_i)^2} = \frac{\sum_i\left( \frac{\Delta w_i}{w_i}\right)^2w^2_i}{\left(\sum_i w_i\right)^2}\leq \max_i\left(\frac{(\Delta w_i)^2}{w_i^2}\right)\frac{\sum w_i^2}{\left(\sum_i w_i\right)^2}\\
\Rightarrow \qquad \frac{\Delta \mathcal{O}}{\mathcal{O}}&\leq\frac{1}{\sqrt{\alpha N}}\max_i{\left(\frac{\Delta w_i}{w_i}\right)}\;,
\end{align}
where we have introduced the Kish effective sample size~\cite{Kish:1964samp} 
\begin{align}
     N_\text{eff} = \frac{\left(\sum_i w_i \right)^2}{\sum_i w_i^2}=\alpha N \qquad \mwith \qquad \alpha\in[0,1]\;.
    \label{eq:Kish}
\end{align}
However, assuming full un-correlation is first unrealistic but also not fully general since one can think that for some part of the phase-space and/or for specific observable those errors will be fully correlated, \ie
\begin{align}
\Delta \mathcal{O} &= \frac{1}{N} \sum_i \Delta w_i\;.
\end{align}
In that case, the relative uncertainty does not vanish anymore but is bounded as
\begin{align}
\frac{\Delta \mathcal{O}}{\mathcal{O}} &= \frac{\sum_i \Delta w_i}{\sum_i w_i} = \leq \frac{\sum_i w_i \, \max_i(\frac{\Delta w_i}{w_i})}{\sum_i w_i}= \max_i\left(\frac{\Delta w_i}{w_i}\right).
\end{align}
This means that even in the worst-case scenario, where all weights are biased in the same direction by, say, X\%, all observables would be biased. But, the relative uncertainty on all observables would remain controlled and would correspond exactly to X\%.

Consequently, the task at hand is to control the uncertainty of the surrogate across the full fiducial phase space. Before introducing our algorithm, we first review other sources of theoretical uncertainty to determine the required accuracy of the surrogate. For leading-order event generation, the dominant source of uncertainty arises from scale variation, which is typically of the order of fifty percent for the normalization and at least ten percent for the shape. A second source of uncertainty comes from the parton distribution functions, corresponding to an error of about three percent (and likely larger in the tails of distributions). Therefore, any numerical error below these scales across the full phase space can be considered acceptable. As a standard benchmark for this paper, we target a one-sigma error (66\% confidence level) of approximately one percent.

As a matter of fact, none of our surrogates achieve this level of precision over the full phase-space.
Therefore, we propose a mixed approach: the surrogate prediction for an event is going to be used only if its estimated relative predicted uncertainty, defined as $\sigma_{\text{nn}}/A_{\text{nn}}$, is smaller than a chosen uncertainty threshold $U_{\text{thr}}$. Events that do not satisfy this requirement are instead evaluated using the exact amplitude, ensuring accuracy over the full phase-space.

%-------------------------------
\begin{figure}[b!]
    \centering
    \includegraphics[width=0.49\textwidth]{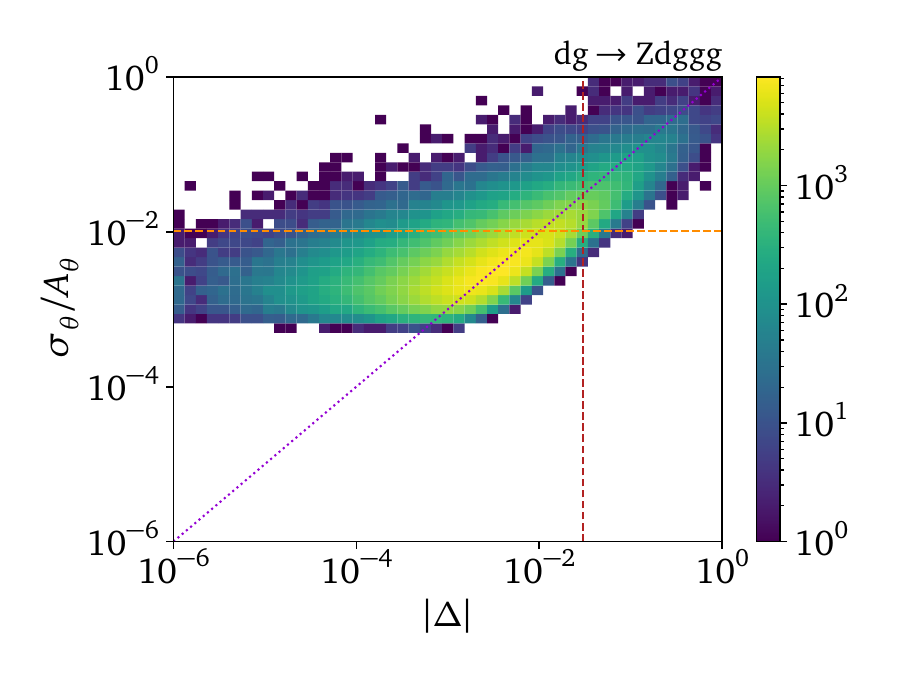}
    \includegraphics[width=0.49\textwidth]{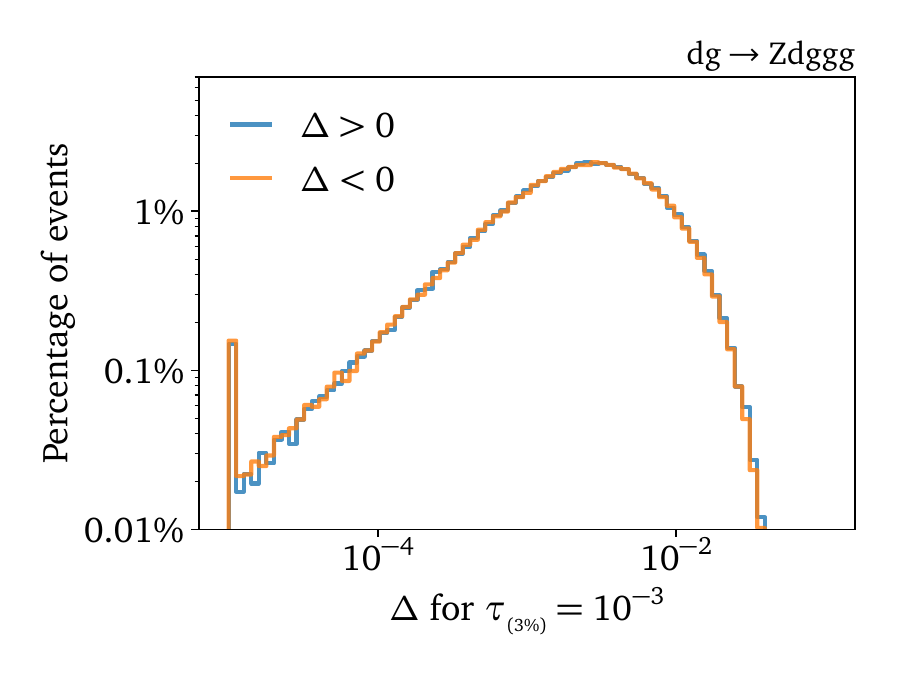}
    \caption{Left: Absolute accuracy vs.\,relative predicted uncertainty for IG1-model on the whole test dataset. The horizontal dashed line corresponds to $U_{\text{thr}}$ for $\tau_{\scriptscriptstyle(3\%)}=10^{-3}$. The vertical dashed line corresponds to the target accuracy $\Delta=3\%$. Right: Accuracy distribution for the events using the surrogates for  $\tau_{\scriptscriptstyle(3\%)}=10^{-3}$ split into negative and positive contribution, for IG1-model on the $\Pd \Pg \to \PZ \Pd \Pg \Pg \Pg$ process.}
    \label{fig:Plot_dg_zd3g_uncert}
\end{figure}
%-------------------------------

%-------------------------------
\begin{figure}[t!]
    \centering
    \includegraphics[width=0.49\textwidth]{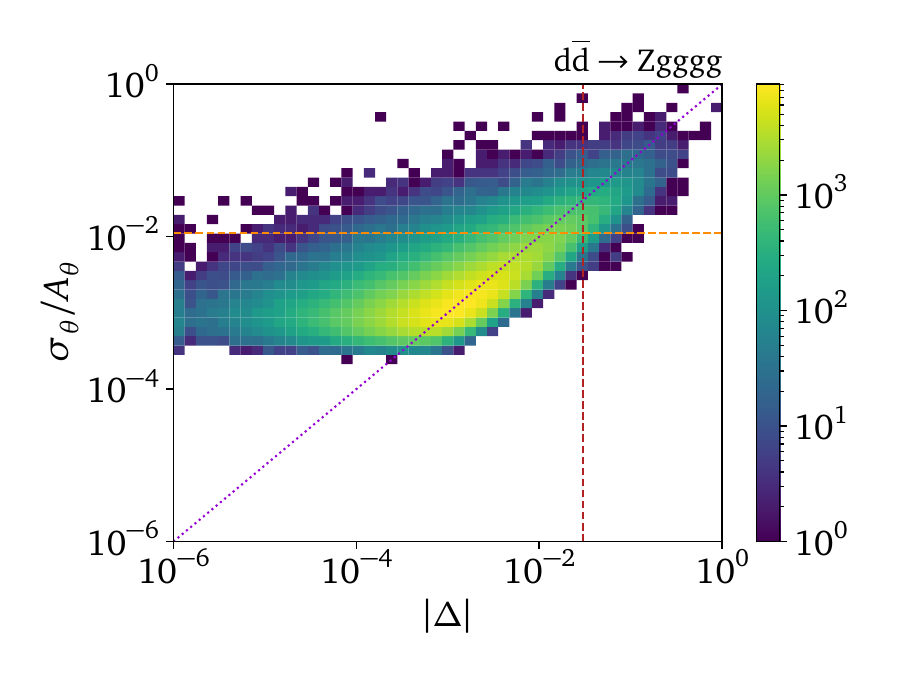}
    \includegraphics[width=0.49\textwidth]{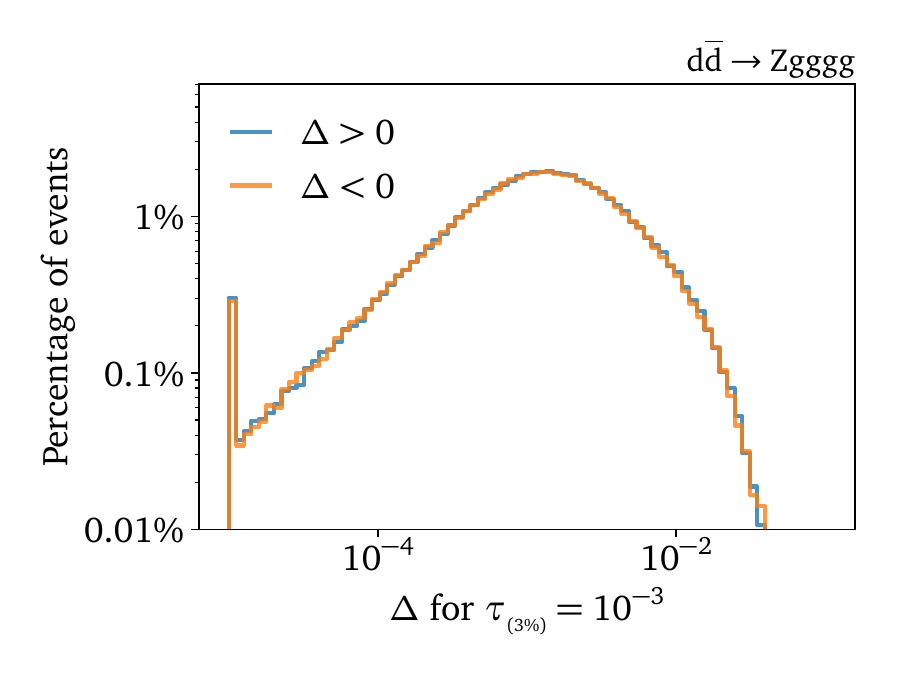}
    \caption{Left: Absolute accuracy vs.\,relative predicted uncertainty for FG1-model on the whole test dataset. The horizontal dashed line corresponds to $U_{\text{thr}}$ for $\tau_{\scriptscriptstyle(3\%)}=10^{-3}$. The vertical dashed line corresponds to the target accuracy $\Delta=3\%$. Right: Accuracy distribution for the events using the surrogates for $\tau_{\scriptscriptstyle(3\%)}=10^{-3}$ split into negative and positive contribution, for FG1-model on the $\Pd \Pdbar \to \PZ \Pg \Pg \Pg \Pg$ process.}
    \label{fig:Plot_ddx_z4g_uncert}
\end{figure}
%-------------------------------

In order to be effective, the method deeply relies on the fact that our estimator of $\sigma_{\theta}$ behaves correctly. To assess that, we display in Fig.~\ref{fig:Plot_dg_zd3g_uncert} and \ref{fig:Plot_ddx_z4g_uncert} (left panel) the correlation between the estimated error from the surrogate and the actual error.
On one hand, we see a clear tendency for $\sigma_{\theta}$ to be over-conservative, especially for very accurate prediction. On the other, we see a quite good correlation between the two quantities. However, one still needs to be careful since some under-estimation can still occur (with normal distribution and therefore for rare event). This is why we only use $\sigma_{\theta}$ as a proxy of the accuracy and not as the real measure. To quantify the additional uncertainty introduced by our approach,  we introduce a tolerance parameter $\tau_{\scriptscriptstyle(x)}$, defined as the fraction of events in the final sample whose accuracy is worse than $x\%$, which means $|\Delta|>x$. 
The tolerance is therefore
\begin{align}
\tau_{\scriptscriptstyle(x)} = \epsilon_U \cdot f^{(x)}_U ,
\end{align}
where $\epsilon_U$ is the fraction of events where $\sigma_{\theta}/A_{\theta} < U_{\text{thr}}$, and $f_U^{(x)}$ is the fraction of those events with accuracy larger than $x$. 

While the choice of acceptable tolerance can vary from one application to another, we have pick for the case of LO phase-space integration a value of $\tau_{\scriptscriptstyle(3\%)}=10^{-3}$.
which, if we assume that everything is Gaussian (which is kind of reasonable) means a 1 sigma error of around 1 percent, our target for LO phase-space generation.
Additionally, one should remember that this 1 sigma error at one percent for any observable is reached in the case where all the error are correlated, which is a very conservative upper-bound. This is confirmed by Fig.~\ref{fig:Plot_dg_zd3g_uncert} and \ref{fig:Plot_ddx_z4g_uncert} (right panel) where we plot independently the events who over-estimate the amplitude from those who under-estimated. For those plots, we have selected only the events passing the respective thresholds, $U_\text{thr} = 0.0102 $ and $U_\text{thr} = 0.0111$, such that we have $\tau_{\scriptscriptstyle(3\%)}=10^{-3}$. Both curves are almost perfectly aligned, suggesting that the error behaves more like in the uncorrelated case than in the correlated one.

Once a tolerance is selected for a given application, we determine—via a simple scan—the optimal value of $U_{\text{thr}}$ required to achieve that tolerance. We then compute the speed-up factor
\begin{align}
f\equiv \frac{t_{\text{MG}}}{t_{\text{tot}}} &=  \frac{t_{\text{MG}}}{t_{\text{surr}} + (1 - \epsilon_U) \cdot t_{\text{MG}}}\;,
\end{align}
where $t_\text{surr}$ is the time to evaluate the surrogate and $t_{\text{MG}}$ is the time to evaluate the true amplitude within \mg.

As usual when quoting any effective gain factor, one need to stress underline hypothesis and context.
The first hypothesis here is that the timing of the computation is highly dominating by the times needed to evaluate such amplitude, allowing us to not take into account the time needed to train such network efficiently, which is likely only valid for the massive HL-LHC simulation of CMS and ATLAS. The second one is that the application under consideration is highly dominated by the time needed to evaluate the amplitude, otherwise one need to apply Amdahl's law \cite{amdahl1967validity} to rescale the speed-up factor down.

This algorithm can naturally be extended to test multiple surrogates sequentially, leveraging the fact that some networks predict certain regions of phase space more accurately than others. For efficiency, we order the surrogates by evaluation cost, and both the speed-up factor and tolerance definition are updated accordingly.
Combining two or more surrogates offers an additional advantage: the ensemble can itself serve as a surrogate. In practice, we first evaluate the faster surrogate; if predicted uncertainty is above our threshold, we then evaluate the second surrogate. The final decision is then based on the ensemble prediction rather than solely on the second surrogate, since the ensemble is at least as accurate and both components have already been computed, the additional cost is virtually zero.

Moreover, using two surrogates allows us to fine-tune the uncertainty thresholds, assigning different thresholds to each step. 
In our study, we found that choosing a smaller threshold for the first surrogate and a slightly larger threshold for the second surrogate (relative to a uniform $U_{\text{thr}}$) improves performance while maintaining the same tolerance. This approach lets us impose stricter constraints on the less accurate network and more relaxed constraints on the more reliable one. 
%To avoid introducing additional variables, in the plots shown below we denote the two-surrogate model using $U_1 = U_{\text{thr}}$ for the first surrogate and $U_2 \sim U_{\text{thr}}\cdot 1.4$ for the second

\subsubsection*{Speed-up for $\Pd \Pg \to \PZ \Pd \Pg \Pg \Pg$}

\begin{figure}[tb!]
    \includegraphics[width=0.49\linewidth]{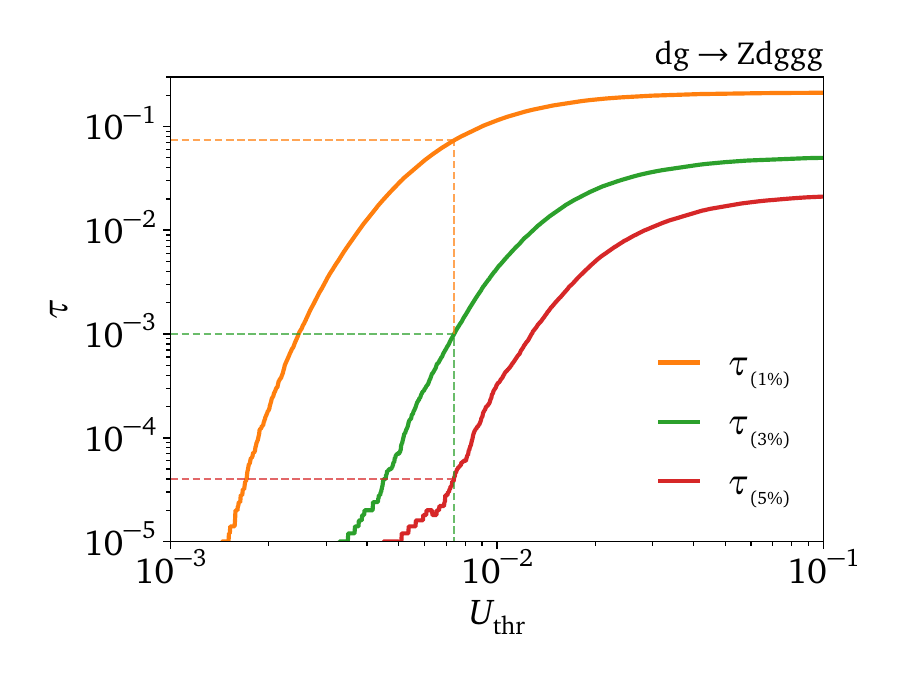}
    \includegraphics[width=0.49\linewidth]{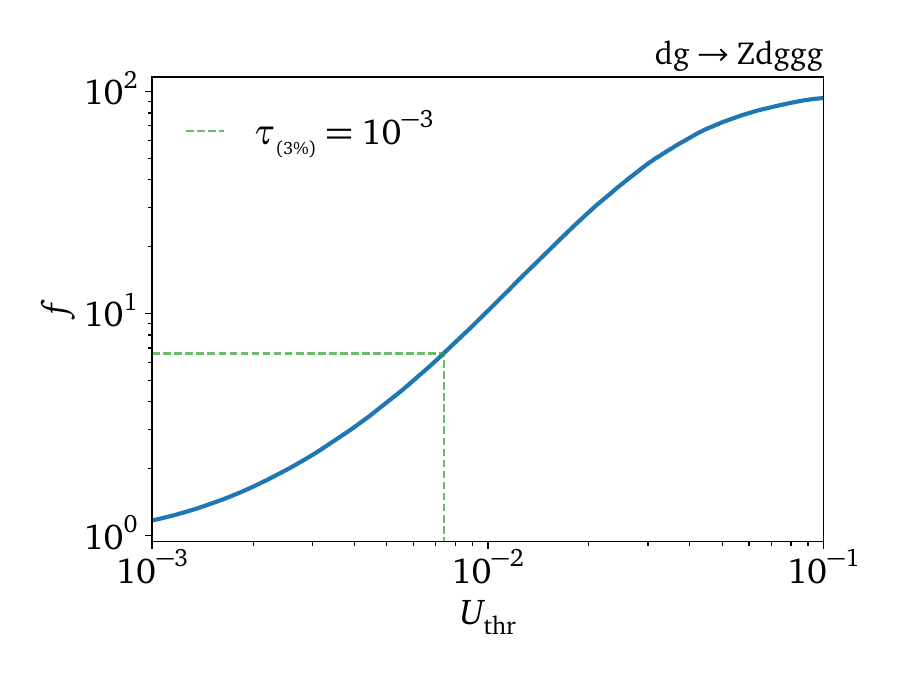}
    \caption{Tolerances (left) and Speed-up factor (right) vs. the uncertainty threshold $U_{\text{thr}}$  for the ensemble [IG1$^2$, FG1] factorization model for the process $\Pd \Pg \to \PZ \Pd \Pg \Pg \Pg$. The plots are obtained using different uncertainty thresholds for the first and second surrogate, such that: $U_1 = U_{\text{thr}}$ and $U_2 = 1.4 \cdot U_{\text{thr}}$} 
    \label{fig:Plot_Speed_FixAcc_dg_zd3g_Speedup}
\end{figure}

In Fig. \ref{fig:Plot_Speed_FixAcc_dg_zd3g_Speedup}, we present, in function of the internal threshold $U_T$, both the value of the tolerance for various target accuracy (left panel) and the achieved speed-up (right panel) for the process $\Pd \Pg \to \PZ \Pd \Pg \Pg \Pg$. For those particular plots, we choose to use the best performing model for this process, but other factorization ansatz have similar behaviour.
On the left panel, the interesting point to note for our benchmark point, corresponding to $\tau_{\scriptscriptstyle(3\%)} = 10^{-3}$ (the green dashed line) is the value of the tolerance for the other threshold. We see that $\tau_{\scriptscriptstyle(1\%)} \sim 7\cdot 10^{-2} $ which is for the 1 percent is actually much better than one would expect given a pure normal distribution of error, which we relate to the fact that our network tend to be over-conservative. 
The $\tau_{\scriptscriptstyle(5\%)} \sim 4 \cdot 10^{-5}$ being more consistent with the theoretical expectation.  This comforts us on the decision to use $\tau_{\scriptscriptstyle(3\%)} = 10^{-3}$ as a reasonable threshold. We will therefore only focus on $\tau_{\scriptscriptstyle3\%}$ afterwards. 
% Two unusual features can be observed in $\tau_{\scriptscriptstyle(5\%)}$: first a change in shape around $7 \cdot 10^{-3}$, and second, a fluctuation of the curve near that value. These behaviors occur only when two consecutive surrogate networks are used and are related to the presence of two moving thresholds. In addition, the statistical sample is limited for these bins.

In the right panel, we show the behaviour of the achieved speed-up. On the left side of the plot we find the region with stringent tolerance constraints, which leads to the rejection of most surrogate-evaluated events. As a consequence, the full amplitude must be computed for the majority of events, and the speed-up remains close to 1. Conversely, the right side of the plot corresponds to the region with looser constraints, where most events are accepted during the first surrogate evaluation. In this regime, the speed-up reaches its maximum value, approaching the ratio between the cost of evaluating the doubly reduced amplitude and that of the full one. 

%---------------------------
%---------------------------

\begin{figure}[b!]
    \includegraphics[width=0.49\linewidth]{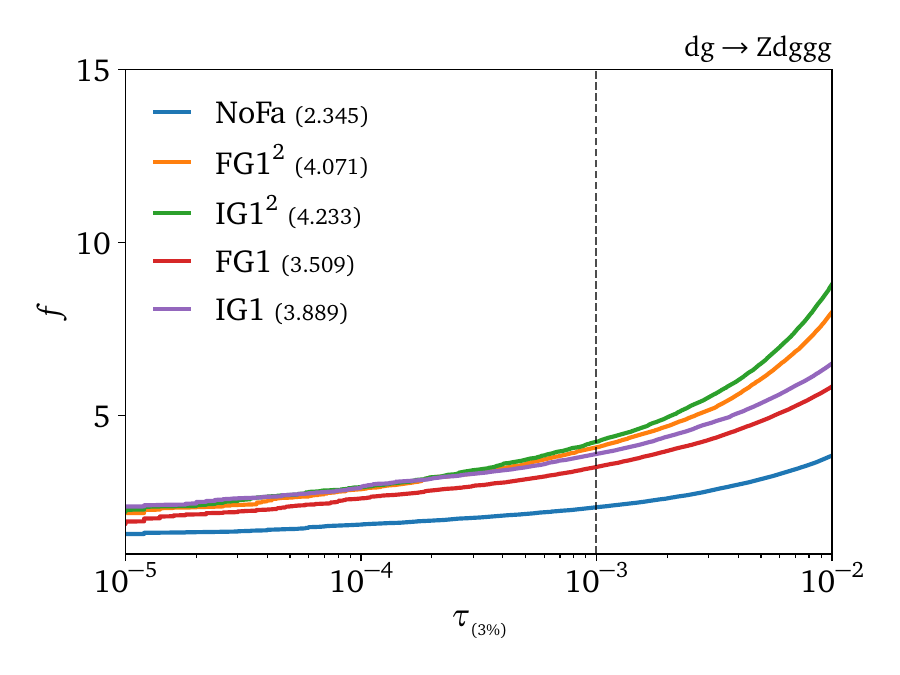}
    \includegraphics[width=0.49\linewidth]{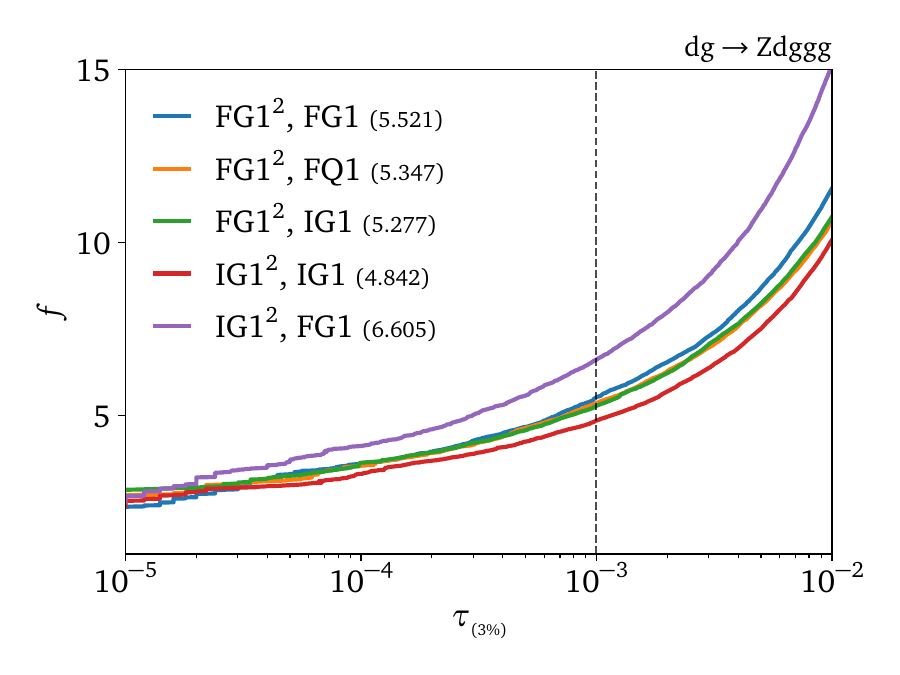}
    \caption{Speed-up factor comparison between different individual factorization models (left), and ensemble factorization models (right) for the process $\Pd \Pg \to \PZ \Pd \Pg \Pg \Pg$. The values in parentheses are the speed-up factors corresponding for $\tau_{\scriptscriptstyle(3\%)} = 10^{-3}$.} 
    \label{fig:Plot_Speed_FixAcc_dg_zd3g_Tol03}
\end{figure}

In Figure~\ref{fig:Plot_Speed_FixAcc_dg_zd3g_Tol03}, we present the final speed-up factor for a given value of the tolerance at $3\%$, enlightening the value of $\tau_{\scriptscriptstyle(3\%)} = 10^{-3}$ as our standard benchmark point. In this plot, we compare the speed-up factors for our different surrogate networks. The results are split into two panels: the left shows single-network configurations, while the right shows ensembles of two surrogate networks (which therefore require two training). It can be seen that, although the no-factorization network is the fastest to evaluate, its effective speed-up factor is quite small due to its limited accuracy—both in predicting the amplitude and in estimating $\sigma_{NN}$). At the same time, the double-factorization model slightly outperforms the single-factorization model, primarily because it is approximately ten times faster to evaluate while still providing reasonably accurate results. Interestingly, this advantage disappears for stricter tolerances  ($\tau_{\scriptscriptstyle(3\%)}\leq3 \cdot 10^{-4}$).

%---------------------------
%---------------------------

Looking  now at the result when using two different surrogates, one can observe different effect due to the subtle interplay between the speed and the accuracy of the surrogates. 
The winning strategy is actually to use the most accurate ensemble of double and single factorization, as shown in Fig.~\ref{fig:Plot_NN_Acc_dg_zd3g_ensemble}. Comparing the winning ensemble to the similar models, we find that this network gains both from having the most accurate first surrogate (IG1$^2$), and from the most accurate ensemble ([IG1$^2$, FG1]), and not just the most accurate second surrogate (IG1). 
We can also build ensemble that use NoFa as a first surrogate. However, the performances of those models are worse compared to ensembles of double and single factorization, thanks to their higher accuracy and more meaningful ensemble prediction.

As can be read on the graph, we reach for our benchmark point a speed-up factor of $6.5$, that can be larger if one allows themselves to be less conservative than us on the allowed threshold. On the other direction, if one wants to be even more conservative of either $\tau_{\scriptscriptstyle(3\%)} = 10^{-4}$ or $\tau_{\scriptscriptstyle(3\%)} = 10^{-5}$, one still get speed-up of roughly 4 and 3 respectively.

\subsubsection*{Speed-up for $\Pd \Pdbar \to \PZ \Pg \Pg \Pg \Pg$}

\begin{figure}[t!]
    \includegraphics[width=0.49\linewidth]{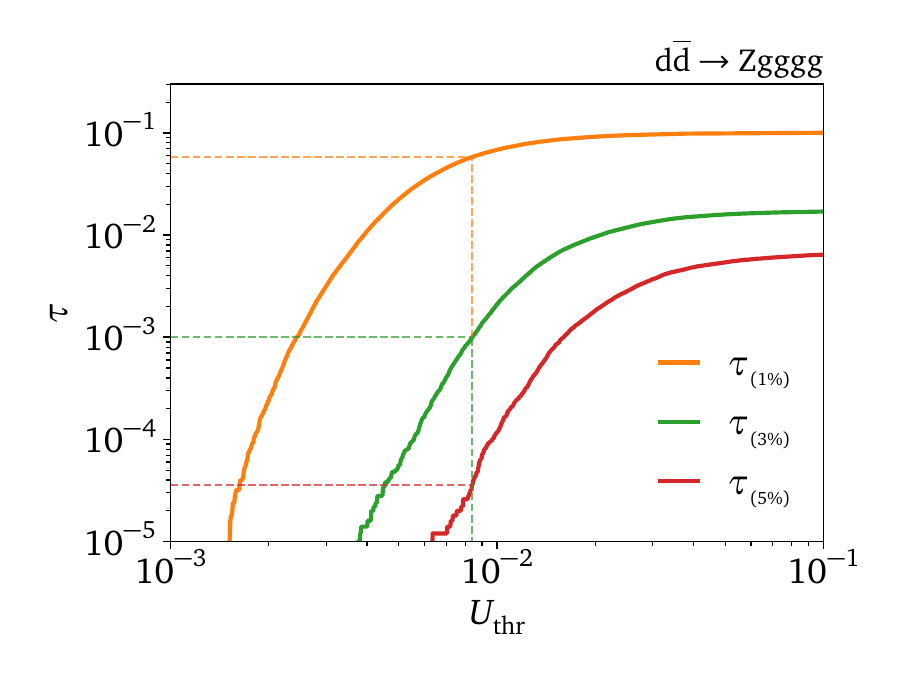}
    \includegraphics[width=0.49\linewidth]{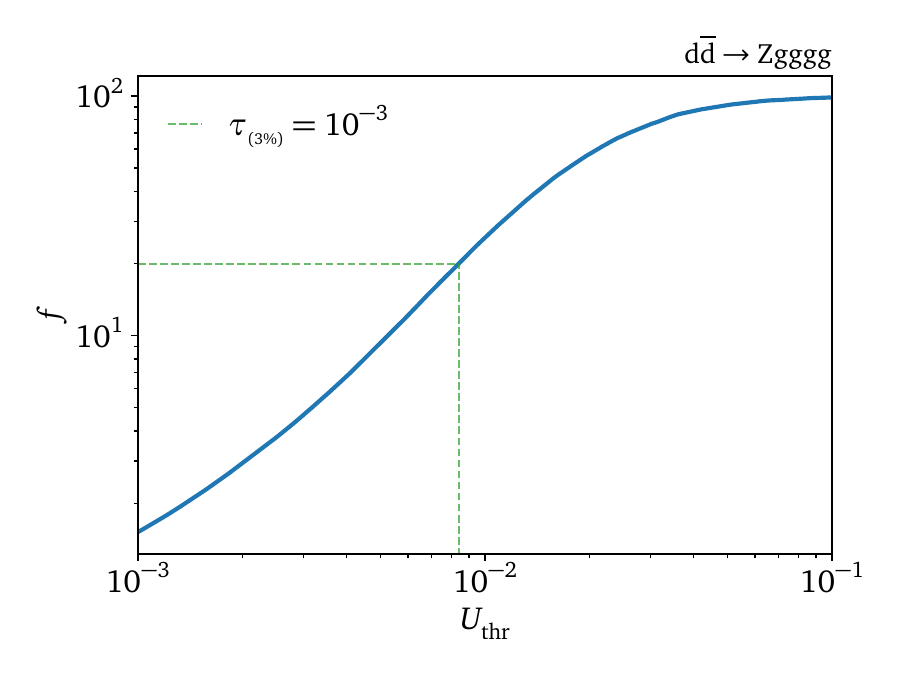}
    \caption{Tolerances (left) and Speed-up factor (right) for the ensemble [FG1$^2$, FG1] factorization model for the process $\Pd \Pdbar \to \PZ \Pg \Pg \Pg \Pg$. 
    The plots are obtained using different uncertainty thresholds for the first and second surrogate, such that: $U_1 = U_{\text{thr}}$ and $U_2 = 1.4 \cdot U_{\text{thr}}$} 
    \label{fig:Plot_Speed_FixAcc_ddx_z4g_Speedup}
\end{figure}

Finally, we present the results for the second process, that is $\Pd \Pdbar \to \PZ \Pg \Pg \Pg \Pg$. 
In Fig.~\ref{fig:Plot_Speed_FixAcc_ddx_z4g_Speedup}, we show the tolerance (left panel) and the achieved speed-up (right panel) as functions of the internal threshold $U_{\text{thr}}$. Comparing the tolerance panel with that of the previous process (Fig.~\ref{fig:Plot_Speed_FixAcc_dg_zd3g_Speedup}), we observe a similar behaviour. However, in this case the threshold $U_{\text{thr}}$ corresponding to a given target tolerance is noticeably larger ($U_{\text{thr}} = 0.0084$ versus $ 0.0073$), reflecting the higher accuracy of the surrogate, which permits looser constraints.

For the same reason, in the right panel, we see that the speed-up increases more rapidly as $U_{\text{thr}}$ grows. Nevertheless, both processes reach the same maximum speed-up, since the evaluation times are identical; the difference lies in the tolerance values at which this plateau is achieved.

%---------------------------
%---------------------------

%---------------------------
\begin{figure}[b!]
    \includegraphics[width=0.49\linewidth]{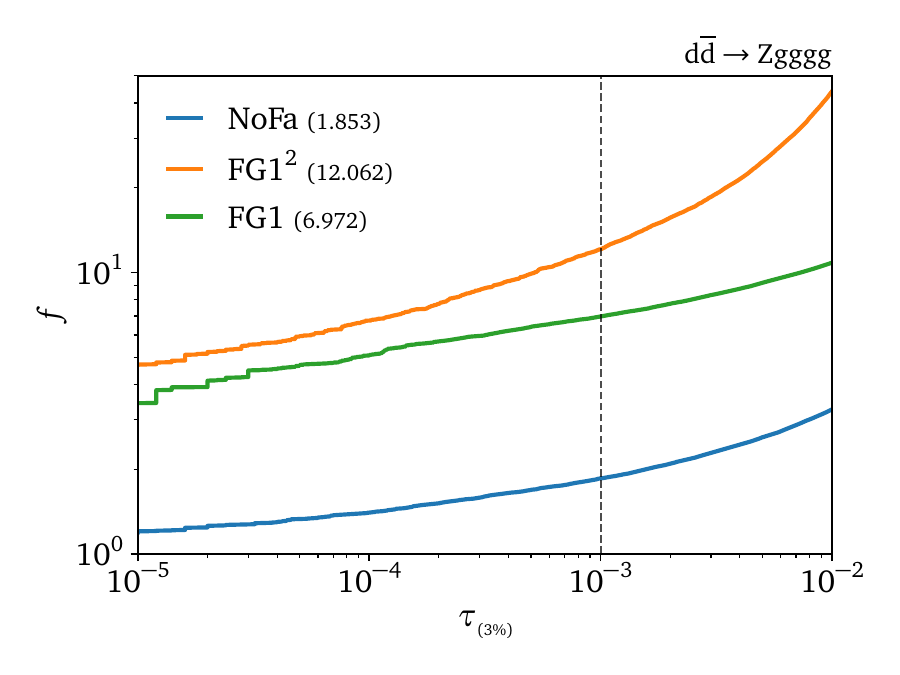}
    \includegraphics[width=0.49\linewidth]{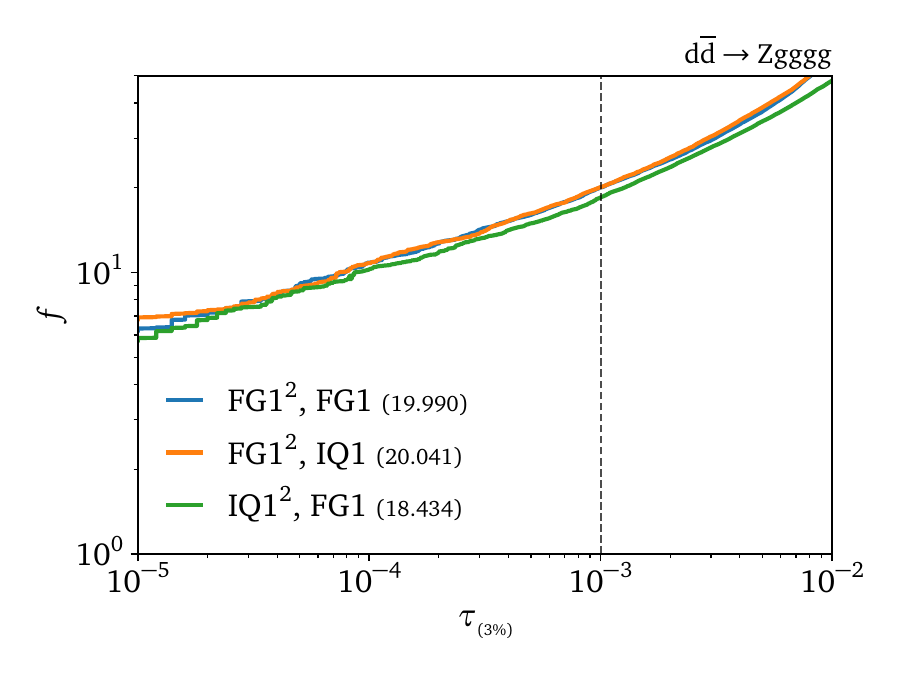}
    \caption{Speed-up factor comparison between different individual factorization models (left), and ensemble factorization models (right) for the process $\Pd \Pdbar \to \PZ \Pg \Pg \Pg \Pg$. The values in parentheses are the speed-up factors corresponding for $\tau_{\scriptscriptstyle(3\%)} = 10^{-3}$.} 
    \label{fig:Plot_Speed_FixAcc_ddx_z4g_Tol03}
\end{figure}
%---------------------------

In Fig.~\ref{fig:Plot_Speed_FixAcc_ddx_z4g_Tol03}, we present the final speed-up factor has a function of the tolerance (at 3 percent), for both individual (left panel) and ensemble networks (right panel).  A clear improvement is observed when transitioning from the NoFa model to a single-factorization surrogate, and subsequently to a double-factorization surrogate. In this case, the double-factorization model is sufficiently accurate to provide nearly a factor-of-two improvement in speed-up compared to the single-factorization model.

For the ensemble models, we find that they achieve similar speed-up factors. As in the other process, the optimal strategy is to use the most accurate double-factorization network as the first surrogate. However, unlike the previous case, we do not observe significant differences between the most accurate ensemble model ([FG1$^2$, IQ1]) and the ensemble combining two highly accurate surrogates ([FG1$^2$, FG1]). 
For our benchmark, we obtain a speed-up factor of approximately 20. As for the previous process, this factor can be higher (and can reach 30) if one allows himself to be less conservative (that speed-up is reached for a still reasonable $\tau_{\scriptscriptstyle(3\%)}=3 \cdot 10^{-3}$). On the opposite direction, the more accurate network allows to still achieve a sizable speed-up of 7, even in the very conservative limit ($\tau_{\scriptscriptstyle(3\%)}=10^{-5}$).

%\clearpage
%%%%%%%%%%%%%%%%%%%%%%%%%%%%%%%%%%%%%%%%%%%%%%%%%%%%%%%%%
\section{Conclusion and outlook}
\label{sec:outlook}

In this paper, we have compared various surrogates to speed-up the evaluation of the amplitude (\ie the matrix-element squared summed/averaged over spin and colour), focusing on the case of LO event generation. For Z+jets processes, we have compared four types of ansatz: starting from a physics agnostic network (dubbed NoFa), which is a simple fully connected neural network, while the three others used a simplified Catani-Seymour approximation to inject additional physics, simplify the function to learn, and ultimately make the prediction more accurate.

Our study demonstrated that, for the non-optimal process we investigated, using physics-based approximations (\ie the Catani-Seymour approximation) led to improved predictions compared to the NoFa network architecture, which directly predicted the logarithm of the amplitude. In contrast, applying the Catani–Seymour approximation twice recursively reduced accuracy to a level comparable to that achieved when the most singular pair was not selected for the splitting kernel.
By training our network with a heteroscedastic loss, the network not only predicted the amplitude but also provided an estimate of its confidence in the prediction. We observed that this confidence was generally pessimistic: the network tended to overestimate the error, predicting low precision even when achieving high precision. Notably, no bias was observed, allowing us to use this estimate conservatively.

First, we leveraged both the error estimates and the correlation matrix to combine predictions from different networks. Although the predictions were highly correlated, each network learned different regions of phase space, yielding a tangible gain in precision.
Second, we used surrogates to replace amplitude evaluations when the (estimated) surrogate accuracy was sufficiently precise. This mixed approach allowed full control over the induced error for any observable computed from LO samples. Considering other sources of uncertainty (PDF and scale variations), we deemed an additional error at the percent level acceptable for all observables and proposed a simple, conservative algorithm to ensure surrogate usage does not compromise predictions and remains below other sources of uncertainty.

Our results revealed a complex trade-off between approximation speed and accuracy. The optimal strategy was neither the fastest (NoFa) nor the most accurate (FG1), but rather the double-factorization network. This network, approximately 100 times faster than a full QFT computation, offers a substantial speed-up that compensates for its sub-optimal accuracy.
For the conservative benchmark we selected (which correspond to an conservative additional statistical error of one percent), we achieved a speed-up factor of 6.5 for the least accurate process ($\Pd \Pg \to \PZ \Pd \Pg \Pg \Pg$) and an impressive factor 20 for the process ($\Pd \Pdbar \to \PZ \Pg \Pg \Pg \Pg$).

Consequently, this work highlights the importance of predicting not only the amplitude but also its associated error. This capability is essential for improving surrogate accuracy and for practical applications, as it enables control over the additional numerical uncertainty introduced by surrogates. Finally, we emphasize that the goal of surrogates should not be to achieve maximal accuracy (which cannot exceed the exact computation), but to maximize speed-up for a given target accuracy, determined by the problem at hand. Thus, network speed is as critical as the prediction precision and the prediction of the error itself.

This work serves as a proof of concept. Future efforts will focus on refactoring the approach to make these surrogates accessible to users, enabling them to assess the required sample size to achieve effective gains, even when accounting for training time. Additionally, it will be important to simplify network training, accelerate evaluation, and improve the reliability of error estimation to achieve even higher speed-up factors.

%%%%%%%%%%%%%%%%%%%%%%%%%%%%%%%%%%%%%%%%%%%%%%%%%%%%%%%%%
\section*{Acknowledgements}
We would like to thank Luigi Favaro, Tilman Plehn, Jesse Thaler, Víctor Bresó-Pla and Alessandra Fanfani for the useful discussions. 
Our research is funded by FRS-FNRS (Belgian National Scientific Research Fund) IISN
projects 4.4503.16 (MaxLHC). This
article/publication is based upon work from COST Action CA24146 and CA22130, supported by COST (European
Cooperation in Science and Technology). Computational resources have been provided by the super-
computing facilities of the Université catholique de Louvain (CISM/UCL) and the Consortium des
Équipements de Calcul Intensif en Fédération Wallonie Bruxelles (CÉCI) funded by the Fond de la
Recherche Scientifique de Belgique (F.R.S.-FNRS) under convention 2.5020.11 and by the Walloon
Region.

\clearpage
%%%%%%%%%%%%%%%%%%%%%%%%%%%%%%%%%%%%%%%%%%%%%%%%%%%%%%%%%
\appendix

\section{Catani-Seymour factorization formulas}
\label{sec:CS_formulas}

\subsection*{Catani-Seymour factorization formulas for final state radiation}
Given a process with $n$ particles in the final state, in the singular limit of a final state radiation $p_i \, p_j \to 0$, we can express the amplitudes as~\cite{Catani:1996vz}
\begin{align}
    \langle |\mathcal{M}_{n}|^2\rangle = \sum_{k \neq i,j} \mathcal{D}_{ij, k} + \mathcal{O}(p_i \, p_j)
    \label{eq:CS_dipoles}
\end{align}
where $\mathcal{O}(p_i \, p_j)$ represents non-singular terms in $p_i \, p_j \to 0$, $k$ is a final state spectator and  $\mathcal{D}_{ij,k}$ represents a dipole contribution of the form
\begin{align}
    \mathcal{D}_{ij, k} = -\frac{1}{2 p_i \, p_j} \; 
    \left\langle \mathcal{M}_{n-1}\Bigg|\frac{\mathcal{T}_k \cdot \mathcal{T}_{ij}}{\mathcal{T}_{ij}^2} \mathcal{V}_{ij,k}\Bigg|\mathcal{M}_{n-1} \right\rangle
\end{align}
where the $\mathcal{M}_{n-1}$ is the tree-level amplitude defined on the set of reduced momenta \\
$\left\{ (p_a, p_b \to p_1, \dots, \tilde{p}_{ij}, \tilde{p}_k, \dots, p_{n-1}) \right\}$. $\mathcal{T}_K$ and $\mathcal{T}_{ij}$ are the colour charges of the emitter and spectator, and $\mathcal{V}_{ij,k}$ are splitting matrices in the helicity space of the emitter. 
The splitting matrices, and the reduced momenta, depend on the radiation variables
\begin{align}
y_{ij,k} = \frac{p_ip_j}{p_jp_i + p_jp_k + p_ip_k} \qquad \mand \qquad
z_{ij,k} = \frac{p_ip_k}{p_ip_k + p_jp_k}
\label{eq:rad_variables}
\end{align}
The momenta of the emitter $\tilde{p}_{ij}$ and the spectator $\tilde{p}_k$ of the reduced process are then given by
\begin{align}
\label{eq:cs_remap}
\tilde{p}_{ij} = p_i + p_j - \frac{y_{ij,k}}{1 - y_{ij,k}} p_k \qquad \mand \qquad \tilde{p}_k = \frac{1}{1 - y_{ij,k}} \, p_k\;.
\end{align}
For a quark (or anti-quark) splitting into a quark (or anti-quark) and a gluon, $q_f \to q_f + g_f$, we have
\begin{align}
    \langle s| \mathcal{V}_{q_i g_j,k}) | s^\prime \rangle  = 8 \pi \alpha C_F \, \left[ \frac{2}{1 - z_{ij,k} \, (1 - y_{ij,k})} - (1 + z_{ij,k}) - \epsilon (1 - z_{ij, k}) \right ] 
    \delta^{s s^\prime}
\end{align}
where $s, s^\prime$ are the spin indices of the fermion $\tilde{q}_{ij}$, $\alpha$ is the strong coupling constant and $C_F=\frac{4}{3}$, and $\epsilon$ is a dimensional regularization parameter such that $d = 4 - 2\epsilon$. 
For a gluon splitting into a pair of gluons, $g_f \to g_f + g_f$, we have
\begin{align}
\begin{split}
\langle \mu | \mathcal{V}_{g_ig_j,k} | \nu \rangle  =  16 &\pi \alpha C_A \, \Bigg[ 
-g^{\mu \nu} \left(\frac{1}{1 - z_{ij,k} \, (1 - y_{ij,k})} + \frac{1}{1 - (1 - z_{ij,k}) \, (1 - y_{ij,k})}  - 2 \right) \\
& + (1 - \epsilon) \frac{1}{p_i \, p_j} 
\left(z_{ij, k} p_i^\mu - (1 - z_{ij, k}) p_j^\mu \right)
\left(z_{ij, k} p_i^\nu - (1 - z_{ij, k}) p_j^\nu \right) \Bigg]\;,
\end{split}
\end{align}
where $C_A=3$.
For a gluon splitting into a pair of quark and anti-quark, $g_f \to q_f + \bar{q}_f$, we have
\begin{align}
    \langle \mu |\mathcal{V}_{q_i\bar{q}_j,k} | \nu \rangle  = 8 \pi \alpha T_R \; 
    \left[ -g^{\mu \nu} - \frac{2}{p_i \, p_j} 
    \left(z_{ij, k} p_i^\mu - (1 - z_{ij, k}) p_j^\mu \right)
    \left(z_{ij, k} p_i^\nu - (1 - z_{ij, k}) p_j^\nu \right) \right] ,
\end{align}
where $T_R=\frac{1}{2}$.

%%%%%%%%%%%%%%%%%%%%%%%%%%%%%%%%%%%%%%%%%%%%%%%%%%%%
\subsection*{Catani-Seymour factorization formulae for initial state radiation}

In the singular limit of an initial state radiation $p_a \, p_i \to 0$, we can express the amplitude as ~\cite{Catani:1996vz}
\begin{align}
    \langle |\mathcal{M}_{n}|^2\rangle = \sum_{k \neq i} \mathcal{D}_{ai, k} + \mathcal{O}(p_a \, p_i)
\end{align}
where $\mathcal{D}_{ai,k}$ represents a dipole contribution of the form
\begin{align}
    \mathcal{D}_{ij, k} = -\frac{1}{2 p_a \, p_i} \; \frac{1}{x_{ai,k}} \;  
    \langle \mathcal{M}_{n-1}|\frac{\mathcal{T}_k \cdot \mathcal{T}_{ai}}{\mathcal{T}_{ai}^2} \mathcal{V}_{ai,k}|\mathcal{M}_{n-1} \rangle
\end{align}

In this case, the momenta are redefined as: 
\begin{align}
(p_a, p_b \to p_1, \dots, p_i, p_k, \dots, p_{n+1})
\Longrightarrow
\left\{ (\tilde{p}_a, p_b \to p_1, \dots, \tilde{p}_k, \dots, p_{n}) \right\}_{(a,i,k)}
\end{align}
The associated radiation variables are given by
\begin{align}
x_{ai,k} = \frac{p_kp_a + p_ip_a - p_ip_k}{(p_k + p_i)\,p_a}
\qquad \mand \qquad
u_{ai,k} = \frac{p_i p_a}{p_i p_a + p_k  p_a}\;.
\end{align}
The momenta of the emitter $\tilde{p}_a$ and the spectator $\tilde{p}_k$ in the reduced process are given by
\begin{align}
\tilde{p}_a = x_{ai,k} \, p_a
\qquad \mand \qquad
\tilde{p}_k = p_k + p_i - (1 - x_{ai,k})\,p_a\;.
\end{align}
For a quark (or anti-quark) splitting into a quark (or anti-quark) and a gluon, $q_i \to q_i + g_f$, we have
\begin{align}
    \langle s | \mathcal{V}_{q_a g_i,k} | s^\prime \rangle  = 8 \pi \alpha C_F \, 
    \left[ \frac{2}{1 - x_{ai,k} + u_{ai,k}} - (1 + x_{ai,k}) - \epsilon(1 - x_{ai,k}) \right] 
    \delta^{s s^\prime}
\end{align}
For a gluon splitting into a pair of gluons, $g_i \to g_i + g_f$, the splitting function is
\begin{align}
\begin{split}
\langle \mu | \mathcal{V}_{g_a g_i,k} | \nu \rangle  &= 16 \pi \alpha C_A \, 
\Bigg[ 
-g^{\mu \nu} 
\bigg(\frac{1}{1 - x_{ai,k} + u_{ai,k}} - 1 + x_{ai,k} \, (1 - x_{ai,k}) \bigg)\\
& + (1 - \epsilon) \frac{1 - x_{ai,k}}{x_{ai,k}} \frac{u_{ai,k} (1 - u_{ai,k})}{p_i \, p_j} \bigg( \frac{p_i^\mu}{u_{ai,k}} - \frac{p_k^\mu}{1 - u_{ai,k}} \bigg) 
\bigg( \frac{p_i^\nu}{u_{ai,k}} - \frac{p_k^\nu}{1 - u_{ai,k}} \bigg)\Bigg] 
\end{split}
\end{align}

\subsection*{Factorization ansatz for final-state radiation}
In our factorization ansatz, we approximate the Catani–Seymour dipole by replacing the spin-correlation tensor with its spin-contracted scalar analogue 
\begin{align}
    -\frac{1}{2 p_i \, p_j} \langle \mu | \mathcal{V}_{ij,k} | \nu \rangle \longrightarrow F_{ij,k}
\end{align}
thereby removing the explicit helicity structure. This eliminates off-diagonal spin correlations and yields a purely scalar splitting kernel.
The colour operator $\mathcal{T}_k \mathcal{T}_{ij}$ is replaced by an effective scalar colour factor $C_{ij,k}$, corresponding to the leading colour approximation.

After summing the colour–spin–correlated Born matrix element over spin and colour indices, the dipole contribution reduces to the simplified form 
\begin{align}
    D_{ij,k} = \langle |\mathcal{M}_{n-1}|^2\rangle \cdot C_{ij,k} \cdot F_{ij, k}
\end{align}
In our network we implement only a single dipole factorization and do not sum over all possible spectators, avoiding the sum in Eq.\eqref{eq:CS_dipoles}. Furthermore, since the ansatz is ultimately multiplied by a neural-network–predicted correction factor, we neglect the explicit colour factor $C_{ij,k}$ allowing the network to absorb this dependence. In this way, our approximation for the full squared amplitude becomes:
\begin{align}
    \langle |\mathcal{M}_{n}|^2\rangle \longrightarrow \langle |\mathcal{M}_{n-1}|^2\rangle \cdot F_{ij, k}
\end{align}

The splitting function $F_{ij,k}^r$ depends on the specific type of radiation. For a quark (or anti-quark) splitting into a quark (or anti-quark) and a gluon, $q_f \to q_f  g_f$, we have
\begin{align}
    F_{ij,k}^{q_f \to q_f g_f}  = \frac{4 \pi \alpha C_F}{p_i \, p_j} \, \left[ \frac{2}{1 - z_{ij,k} \, (1 - y_{ij,k})} - 1 + z_{ij,k} \right ] 
\end{align}
where $\alpha$ is the strong coupling constant and $C_F=\frac{4}{3}$.
%is Casimir operators of the fundamental representation of the $SU(3)$. 
For a gluon splitting into a pair of gluons, $g_f \to g_f  g_f$, we have:
\begin{align}
    F_{ij,k}^{g_f \to g_f g_f}  =  \frac{8 \pi \alpha C_A}{p_i \, p_j} \, \left[ \frac{1}{1 - z_{ij,k} \, (1 - y_{ij,k})} + \frac{1}{1 - (1 - z_{ij,k}) \, (1 - y_{ij,k})}  - 2 \right]\;,
\end{align}
where $C_A=3$.
For a gluon splitting into a pair of quark and anti-quark, $g_f \to q_f  \bar{q}_f$, we have
\begin{align}
    F_{ij,k}^{g_f \to q_f \bar{q}_f}  = \frac{4 \pi \alpha T_R}{p_i p_j} \;,
\end{align}
where $T_R=\frac{1}{2}$.

\subsection*{Factorization ansatz for initial-state radiation}

In the case of initial state radiation, the momenta are redefined as: 
\begin{align}
(p_a, p_b \to p_1, \dots, p_i, p_k, \dots, p_{n+1})
\Longrightarrow
\left\{ (\tilde{p}_a, p_b \to p_1, \dots, \tilde{p}_k, \dots, p_{n}) \right\}_{(a,k)}
\end{align}
The associated radiation variables are given by
\begin{align}
x_{ai,k} = \frac{p_kp_a + p_ip_a - p_ip_k}{(p_k + p_i)\,p_a}
\qquad \mand \qquad
u_{ai,k} = \frac{p_i p_a}{p_i p_a + p_k  p_a}\;.
\end{align}
The momenta of the emitter $\tilde{p}_a$ and the spectator $\tilde{p}_k$ in the reduced process are given by
\begin{align}
\tilde{p}_a = x_{ai,k} \, p_a
\qquad \mand \qquad
\tilde{p}_k = p_k + p_i - (1 - x_{ai,k})\,p_a\;.
\end{align}
For a quark (or anti-quark) splitting into a quark (or anti-quark) and a gluon, $q_i \to q_i  g_f$, we have
\begin{align}
    F_{ai,k}^{q_i \to q_i  g_f}  = \frac{8 \pi \alpha C_F}{x_{ai,k}p_ip_i} \, \left[ \frac{2}{1 - x_{ai,k} + u_{ai,k}} - 1 - x_{ai,k} \right] 
\end{align}
For a gluon splitting into a pair of gluons, $q_i \to q_i  g_f$, the splitting function is
\begin{align}
    F_{ai,k}^{q_i \to q_i  g_f}  = \frac{8 \pi \alpha C_A}{x_{ai,k}p_ip_i} \, \left[ \frac{1}{1 - x_{ai,k} + u_{ai,k}} - 1 + x_{ai,k} \, (1 - x_{ai,k}) \right ] 
\end{align}

\clearpage
%%%%%%%%%%%%%%%%%%%%%%%%%%%%%%%%%%%%%%%%%%%%%%%%%%%%%%%%%
\appendix

\clearpage
% use a customized SciPost BibTeX style with nicer arXiv
\bibliography{refs} % use BibTeX library
\nolinenumbers
\end{document}